\documentclass[epj]{svjour}

\usepackage{latexsym}
\usepackage{amssymb}
\usepackage{pdfpages}
\usepackage{graphics}

\begin{document}
\newcommand{\ja}{Jakubassa-Amundsen }
\newcommand{\bfx}{\mbox{\boldmath $x$}}
\newcommand{\bfxs}{\mbox{{\scriptsize \boldmath $x$}}}
\newcommand{\bfq}{\mbox{\boldmath $q$}}
\newcommand{\bfqs}{\mbox{{\scriptsize \boldmath $q$}}}
\newcommand{\bfnabla}{\mbox{\boldmath $\nabla$}}
\newcommand{\bfeps}{\mbox{\boldmath $\epsilon$}}
\newcommand{\bfsigma}{\mbox{\boldmath $\sigma$}}
\newcommand{\bfsigmas}{\mbox{{\scriptsize \boldmath $\sigma$}}}
\newcommand{\bfalpha}{\mbox{\boldmath $\alpha$}}
\newcommand{\bfA}{\mbox{\boldmath $A$}}
\newcommand{\bfe}{\mbox{\boldmath $e$}}
\newcommand{\bfn}{\mbox{\boldmath $n$}}
\newcommand{\bfj}{\mbox{\boldmath $j$}}
\newcommand{\bfJ}{\mbox{\boldmath $J$}}
\newcommand{\bfW}{{\mbox{\boldmath $W$}_{\!\!rad}}}
\newcommand{\bfM}{\mbox{\boldmath $M$}}
\newcommand{\bfP}{\mbox{\boldmath $P$}}
\newcommand{\bfY}{\mbox{\boldmath $Y$}}
\newcommand{\bfI}{\mbox{\boldmath $I$}}
\newcommand{\bfp}{\mbox{\boldmath $p$}}
\newcommand{\bfk}{\mbox{\boldmath $k$}}
\newcommand{\bfks}{\mbox{{\scriptsize \boldmath $k$}}}
\newcommand{\bfs}{\mbox{\boldmath $s$}_0}
\newcommand{\bfv}{\mbox{\boldmath $v$}}
\newcommand{\bfw}{\mbox{\boldmath $w$}}
\newcommand{\bfb}{\mbox{\boldmath $b$}}
\newcommand{\bfxi}{\mbox{\boldmath $\xi$}}
\newcommand{\bfzeta}{\mbox{\boldmath $\zeta$}}
\newcommand{\bfr}{\mbox{\boldmath $r$}}
\newcommand{\bfrs}{\mbox{{\scriptsize \boldmath $r$}}}
\newcommand{\bfy}{\mbox{\boldmath $y$}}
\newcommand{\bfys}{\mbox{{\scriptsize \boldmath $y$}}}
\newcommand{\bfps}{\mbox{{\scriptsize \boldmath $p$}}}

\renewcommand{\theequation}{\thesection.\arabic{equation}}
\renewcommand{\thesection}{\arabic{section}}

\title{\Large\bf On the excitation of the $2^+_1$ state in $^{12}$C in the $(e,e'\gamma)$ reaction }
\titlerunning{On the excitation of the $2^+_1$ state in $^{12}$C in the $(e,e'\gamma)$ reaction }


\author{D.~H.~Jakubassa-Amundsen\inst{1} and V.~Yu.~Ponomarev\inst{2}}

\institute{
Mathematics Institute, University  of Munich, Theresienstrasse 39, 80333 Munich, Germany
\and
Institute of Nuclear Physics, Technical University of Darmstadt, 64289 Darmstadt, Germany
}

\date{Received: date / Revised version: date}


\vspace{1cm}

\abstract{  
The excitation of the carbon $2^+$ state at 4.439 MeV by $70-150$ MeV electron impact and its subsequent decay to the ground state by photon emission
is described within the distorted-wave Born approximation. The transition densities are obtained from the nuclear quasiparticle phonon model.
The photon angular distributions are compared with earlier results and with experiment, including the influence of bremsstrahlung.
Predictions for spin asymmetries in the case of polarized electron impact are also made.
\PACS{25.30Bf, 24.70+s, 34.80.Nz}
}
 
\maketitle

\section{Introduction}

With the advent of modern accelerators and efficient spin-polarized electron sources, such as the Darmstadt facility
S-DALINAC, coincidence measurements between electrons scattered inelastically from nuclei and decay photons are feasible with high accuracy.
This has stimulated the theoretical reinvestigation of the lowest quadrupole excitation of $^{12}$C by electron impact
which had been studied in a pioneer coincidence experiment by Papanicolas et al \cite{Pa85}, followed by a theoretical interpretation by Ravenhall et al \cite{Ra87}.

Nuclear excitation by electron impact is a powerful tool to obtain nuclear structure information \cite{Ub71,HB83}, because only the electromagnetic interaction
between the parti\-ci\-pating particles is involved. 
The nuclear properties enter exclusively into the electric and magnetic transition densities $\varrho_L$ and $J_{L,L\pm1}$.
They can be calculated from nuclear models.

The first theoretical investigation of the coincident nuclear excitation and decay (ExDec) process dates back to Hubbard and Rose \cite{HR66}, 
who employed the plane-wave Born approximation (PWBA).
This theory was subsequently applied to the $2_1^+$ excitation of $^{12}$C \cite{AR67}.
Later, a combination of the distorted-wave Born approximation (DWBA) for the electric transition and the PWBA for the magnetic transition was used \cite{Ra87}.
In these calculations, the nuclear transition densities were taken in the form
of Fourier-Bessel series with coefficients obtained from a fit 
to early measurements of inclusive electron scattering form factors 
(e.g. \cite{Cra66,Fla78}).
It has been demonstrated that the current transition densities $J_{L,L\pm 1}$ 
are very strong for the $^{12}$C nucleus, such that interference phenomena 
between electric and magnetic excitations are already visible at scattering 
angles in the forward hemisphere where the cross sections are large.
Such interference effects, augmented in coincidence experiments, are 
particularly sensitive to details of the nuclear structure.

A competitive process to ExDec is the emission of bremsstrahlung, which contributes coherently to the photon emission from nuclear decay to the ground state \cite{HR66}.
Bremsstrahlung calculations at high collision energies are usually performed within the relativistic PWBA.
Since $^{12}$C is a spin-zero nucleus, only potential scattering has to be taken into account \cite{BH34}.
For the radiation of photons with small frequencies as compared to the collision energy (i.e. for low momentum transfer), it was shown by Bethe and Maximon \cite{BM54} that the PWBA, as limiting case of the Sommerfeld-Maue theory, is an appropriate theory, irrespective of the nuclear charge.

Bremsstrahlung may have a considerable influence on the angular distribution of the emitted photons, and was already taken into account in \cite{AR67}, however not in the $^{12}$C investigation by Ravenhall et al \cite{Ra87}.
More recently, when studying the $2_1^+$ and $2_2^+$ excitations in the $(e,e'\gamma)^{92}$Zr reaction \cite{JP17}, it was shown that the contribution of
bremsstrah\-lung to the detected photons
depends not only on the scattering angle, but also on the resolution of the photon detector, which in general is much poorer than the line width of the decay photon.

Within a new campaign of the coincidence experiments in the $(e,e'\gamma)$
reaction at S-DALINAC \cite{SFB}, it is planned to revisit the previous
measurements in \cite{Pa85} to test the set-up.
In the present paper, we extend the theoretical analysis for this experiment.
We employ a full DWBA prescription of the ExDec process 
and add bremsstrahlung coherently.
This guarantees a consistent representation of all interference effects, which are absent in PWBA.
We will use the charge and current transition densities from \cite{Ra87}
and also the ones from the random phase approximation of
the quasiparticle phonon model (QPM \cite{So92,IP12})
to discuss the nuclear structure effects on the ExDec cross sections.
Section 2 provides the differential cross section results.
We also compare with results where the QPM transition densities are replaced by the ones fitted to experiment.
Section 3 deals with the Sherman function for polarized electrons. A short conclusion is given in section 4. Atomic units ($\hbar=m=e=1$) are used unless indicated otherwise.

\section{Cross section for the ExDec process}

The triply differential cross section for the inelastic scattering of an unpolarized electron with (total) initial energy $E_i$ and final energy $E_f$ into the solid angle $d\Omega_f$ with the simultaneous emission of a 
photon with frequency $\omega$ into the solid angle $d\Omega_k$ is given by \cite{HR66,JP17}
$$\frac{d^3\sigma}{d\omega d\Omega_k d\Omega_f}\;=\;\frac{2\pi^2\omega^2E_iE_fk_f}{k_ic^7\;f_{\rm rec}}\;\sum_{\sigma_i,\sigma_f}$$
\begin{equation}\label{2.1}
\times \;\sum_\lambda \left| M_{fi}^{(1)}\;+\;M_{fi}^{(2)}\;+\;M_{fi}^{\rm brems}\right|^2,
\end{equation}
where $\bfk_i$ and $\bfk_f$ are, respectively, the electron momenta in initial and final state. Here we have assumed that polarization is not observed, such that
 (\ref{2.1}) includes a sum over the photon polarization  $\bfeps_\lambda$ and over the final electron spin projection  $\sigma_f$, in addition to an average over the initial-state spin projection $\sigma_i$.
Furthermore, it is assumed that a spin-zero nucleus  decays into its ground state, so that no further spin degrees of freedom are present.
The factor $f_{\rm rec}$ is due to the kinematical recoil arising from the finite mass of the nucleus.

The amplitude $M_{fi}^{(1)}$ describes the excitation of the nucleus into a quadrupole state $n$ with energy $E_x$, spin $J_n=2$ and magnetic quantum number $M_n$, followed by  radioactive decay according to the decay width $\Gamma_n$,
$$ M_{fi}^{(1)}\;=\;i\;\frac{Z_Tc^2}{4\pi\sqrt{\omega}}\;\frac{1}{\omega -E_x +i \Gamma_n/2}$$
\begin{equation}\label{2.2}
\times\;\sum_{M_n=-J_n}^{J_n} A_{ni}^{\rm exc}(M_n)\;A_{fn}^{\rm dec}(M_n),
\end{equation}
where $Z_T$ is the nuclear charge number and $A_{ni}^{\rm exc}$ and $A_{fn}^{\rm dec}$ are, respectively, the excitation and decay amplitudes as e.g. given in \cite{JP17}.

The second transition amplitude in (\ref{2.1}), $M_{fi}^{(2)}$, describes the reversed process where the photon emission occurs before the nuclear excitation. This process is, however, suppressed by several orders of magnitude and can be disregarded.

The last term in (\ref{2.1}), $M_{fi}^{\rm brems}$, is the contribution from bremsstrahlung photons with the same frequency $\omega$,
\begin{equation}\label{2.3}
M_{fi}^{\rm brems} \;=\;i\;\frac{c}{\sqrt{\omega}}\int d\bfx\;\psi_f^{(\sigma_f)+}(\bfx)\;(\bfalpha \bfeps_\lambda^\ast)\;e^{-i\bfks\bfxs}\;\psi_i^{(\sigma_i)}(\bfx),
\end{equation}
where $\psi_i$ and $\psi_f$ are, respectively, the initial and final scattering states of the electron,
while $\bfk$ is the photon momentum and $\bfalpha$ is a vector of Dirac matrices.
In the PWBA, when $\psi_i$ and $\psi_f$ are expanded in terms of plane waves, the rhs of (\ref{2.3}) has to be multiplied by the Dirac form factor $F_1(q)$, which accounts for the charge distribution of the nucleus 
\cite{GP64,Jaku13}.

\subsection{Nuclear excitation}

The excitation amplitude $A_{ni}^{\rm exc}$ is conventionally calculated with the help of partial-wave expansions \cite{TW68,Ros}. It is composed of the contributions originating from the electric transition density $\varrho_L(x_N)$ and the magnetic transition densities $J_{L,L\pm 1}(x_N)$ with $L=2$.
Their dependence on the nuclear coordinate $x_N$ is displayed in Fig.1, where the magnetization current densities contributing to $J_{23}$ and $J_{21}$ are shown separately.
The transition densities, calculated within the QPM, are presented by solid
lines. They have been calculated within the one-phonon approximation by
adjusting the strength of the residual interaction to reproduce the
experimental value of the $B(E2, g.s. \to 2^+_1)$ = 39.7~e$^2$fm$^4$
\cite{Raman}. 
Notice that the QPM transition densities 
deviate considerably from those provided in \cite{Ra87} 
which are obtained from a Fourier-Bessel fit to scattering experiments.
The $B(E2, g.s. \to 2^+_1)$ value obtained from the integration of the
charge transition density \cite{HB83},
\begin{equation}\label{2.4}
B(E2, g.s. \to 2^+) = 5
\left| \int_0^{\infty} r^{4} \varrho_{2}(r)dr
\right|^2,
\end{equation}
when using $\varrho_2$ from \cite{Ra87},
equals to 42.41 e$^2$fm$^4$, which is 7\% above the
experimental value. 
A stronger peak at the surface of the charge transition density 
\cite{Ra87} is compensated by a slightly stronger tail of the QPM density.

\begin{figure}[t]

\vspace*{-1.3cm}

\includegraphics[width=10cm]{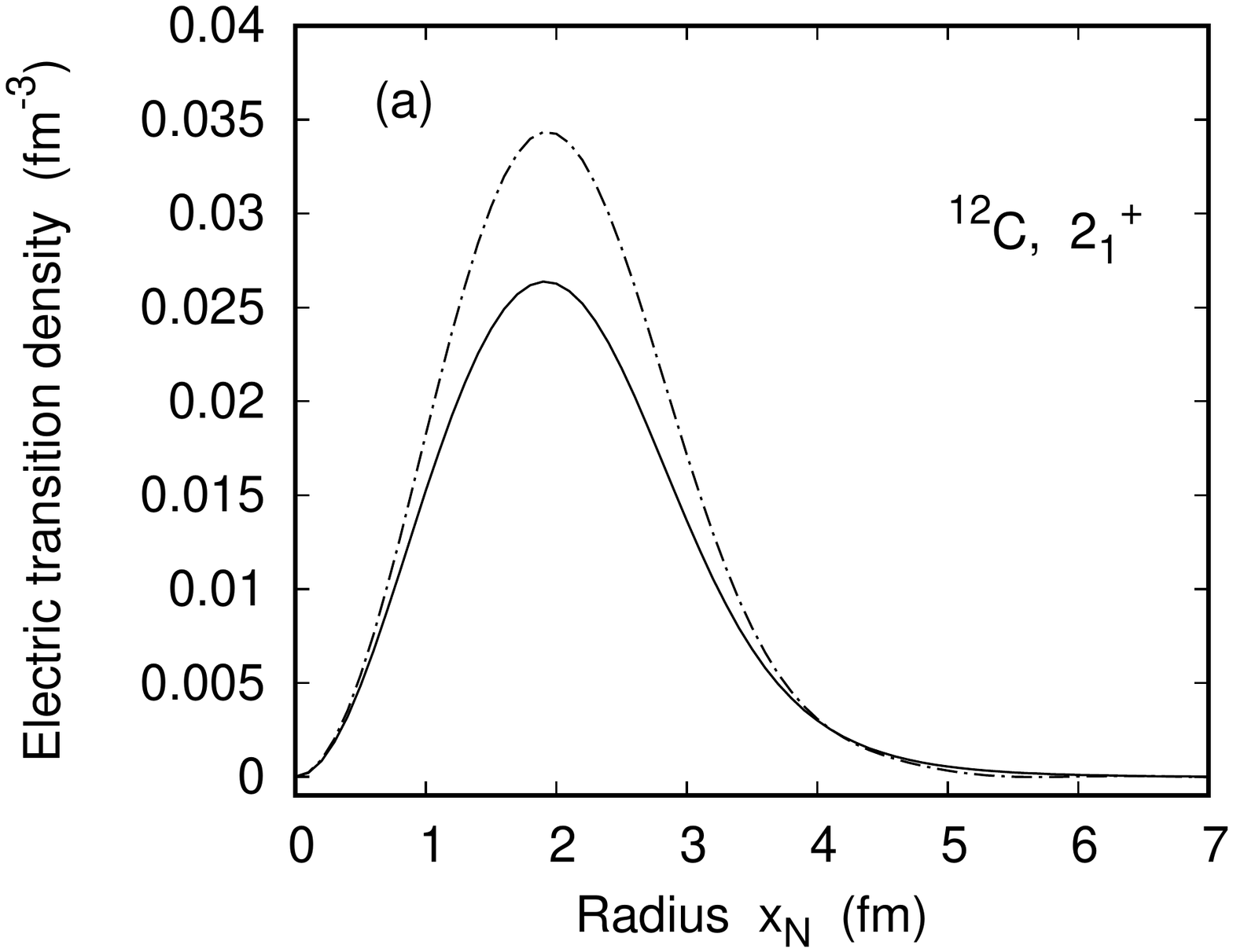}

\vspace*{-2.0cm}

\includegraphics[width=10cm]{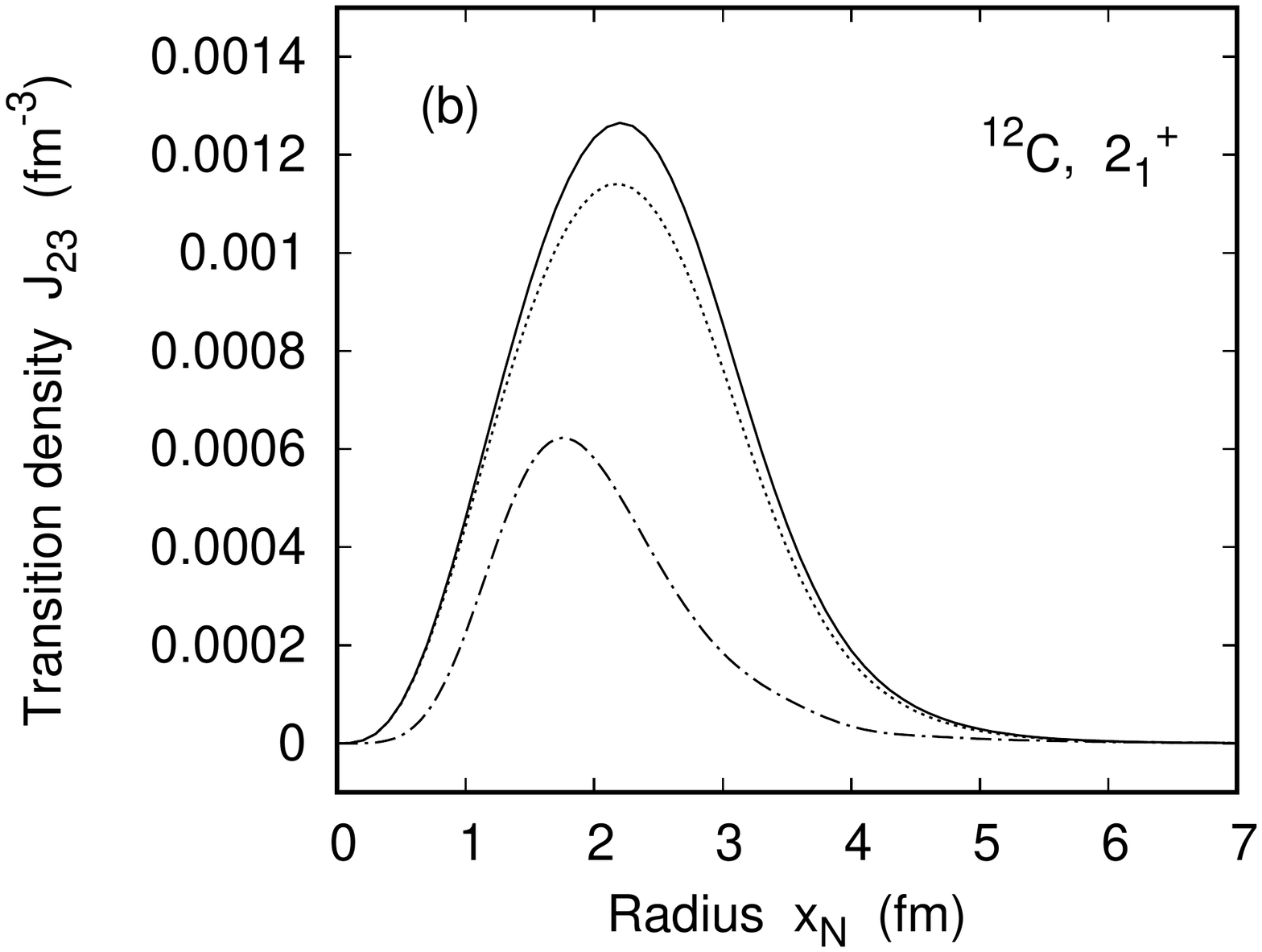}

\vspace*{-1.8cm}

\includegraphics[width=10cm]{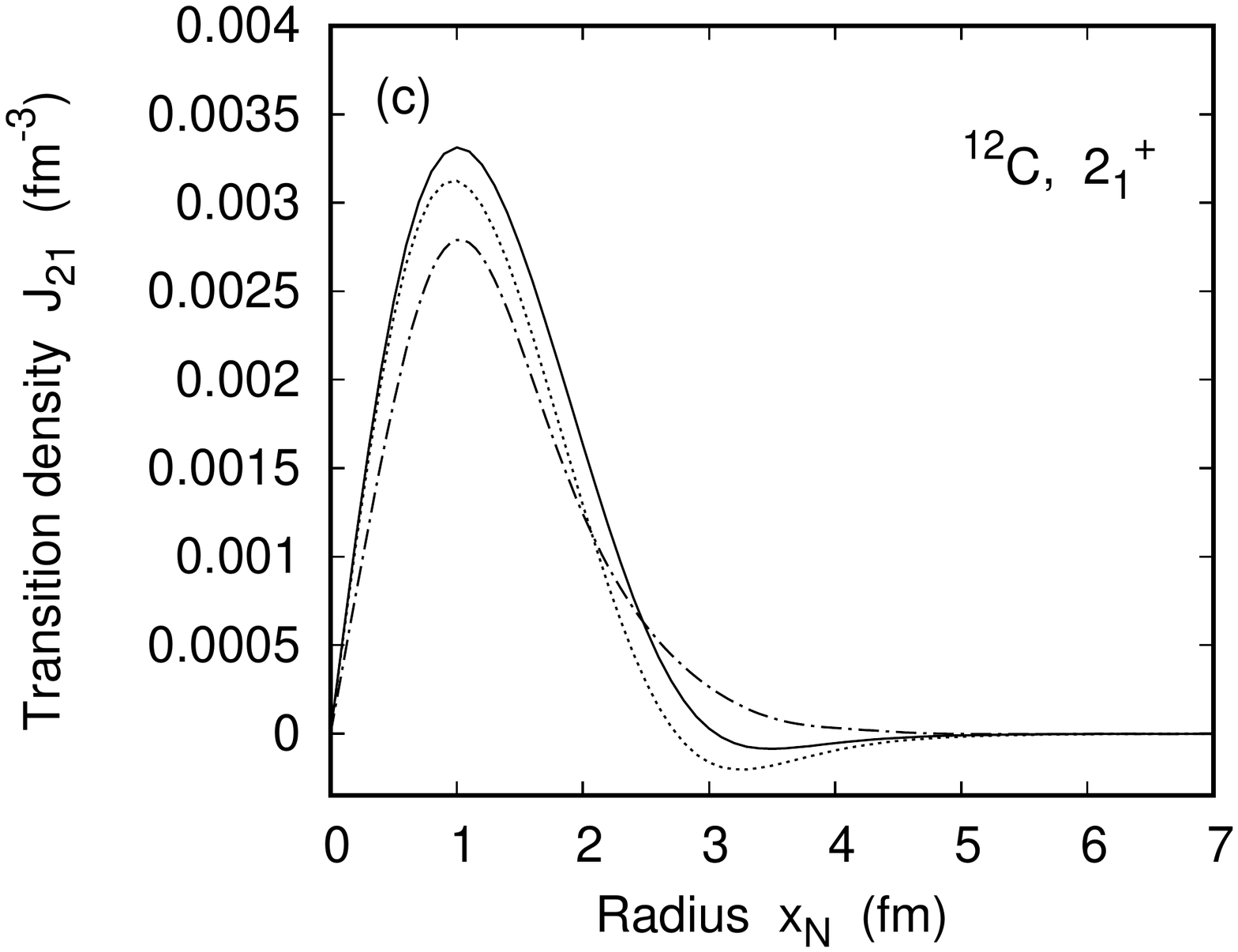}

\vspace*{-.5cm}

\caption
{Transition densities (a) $\varrho_2$, (b) $J_{23}$ and (c) $J_{21}$ for the $2_1^+$ excitation of $^{12}$C at 4.439 MeV as a function of $x_N$. -----------, QPM calculations (consisting in (b) and (c)
 of magnetization and convection currents). $\cdots\cdots$, contribution of the magnetization current to  $J_{23}$ and  $J_{21}$.
 Also shown are the transition densities of Ravenhall et al ($-\cdot - \cdot -$): $\varrho_2$ and $J_{23}$ are taken from \cite{Ra87}, $J_{21}$ is obtained from the continuity equation \cite{HB83}; note, however, the reversed sign of $J_{21}$ as compared to the definition in \cite{HB83}.}
\end{figure}

For quadrupole excitation there are five magnetic sublevels $M_n$, which are populated  with a  probability given by
\begin{equation}\label{2.5}
P(M_n)\;=\;\frac{d\sigma^{\rm exc}/d\Omega_f(M_n)}{(d\sigma^{\rm exc}/d\Omega_f)_{\rm tot}},
\end{equation}
where (see, e.g. \cite{Jaku15})
$$\frac{d\sigma^{\rm exc}}{d\Omega_f}(M_n)\;=\;\frac{2\pi^3 E_iE_fk_f}{k_ic^2\;\tilde{f}_{\rm rec}}\;\sum_{\sigma_i,\sigma_f}$$
\begin{equation}\label{2.6}
\times \;\sum_{M_n'=-2}^2 \left| A_{ni}^{\rm exc}(M_n')\right|^2\;\delta_{M_n,M_n'},
\end{equation}
valid for spin-zero nuclei. The recoil denominator $\tilde{f}_{\rm rec}$ differs from $f_{\rm rec}$ in (\ref{2.1})  due to $E_x$ in the energy balance. The total excitation cross section in the denominator of (\ref{2.5}) results from (\ref{2.6}) with the delta function removed.

The calculation of the exact electronic scattering states by means of the Dirac equation 
is performed with the help of the Fortran code RADIAL by Salvat et al \cite{Sal}.
The nuclear potential of $^{12}$C is generated from the Fourier-Bessel expansion of the ground-state charge distribution \cite{deV}.
The radial integrals in the transition matrix elements are evaluated by means of the complex-plane rotation method (CRM) introduced in \cite{VF70} and applied to electron scattering in \cite{JP16}.

\begin{figure}

\vspace*{-1.3cm}

\includegraphics[width=10cm]{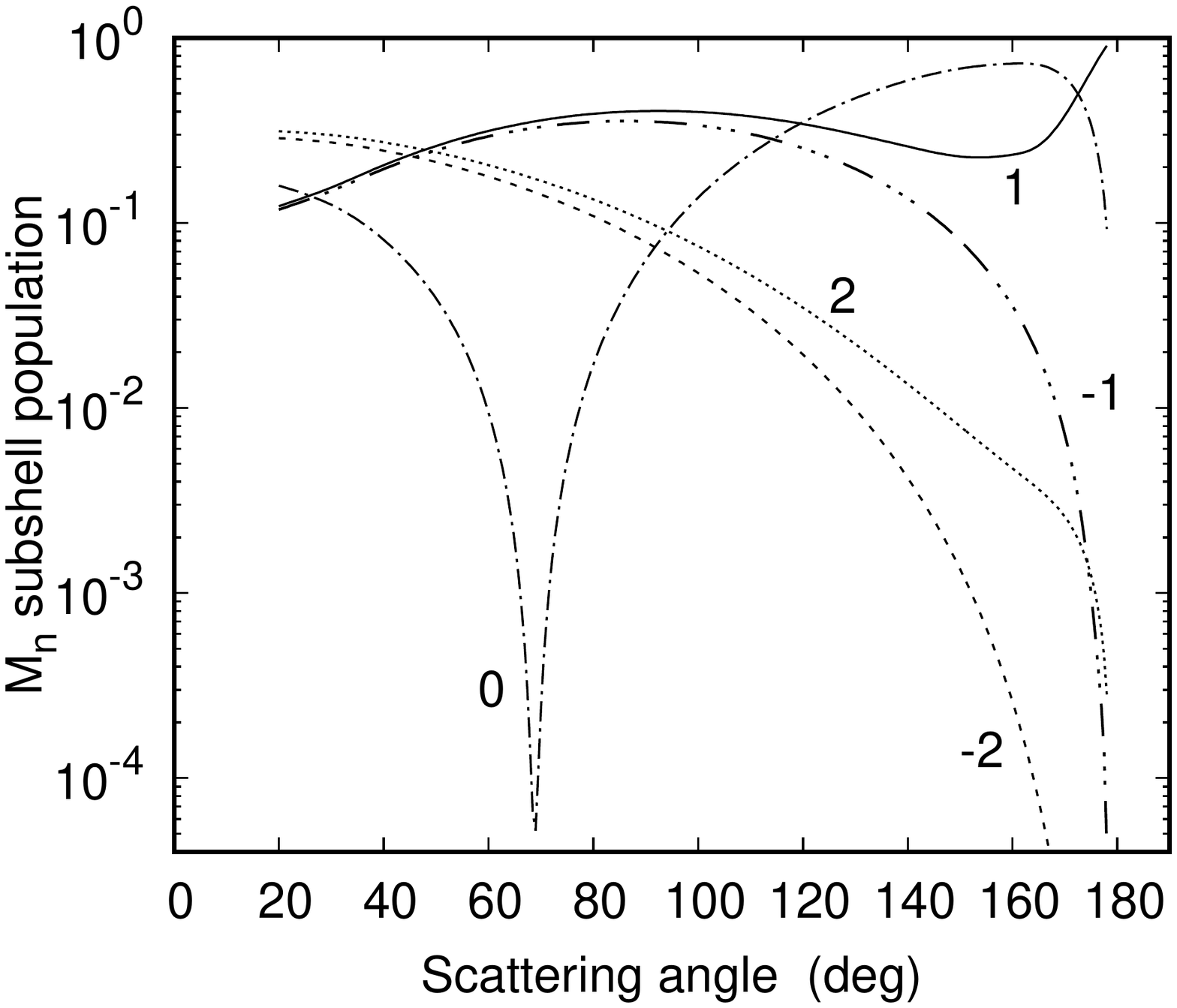}

\vspace*{-0.5cm}

\caption
{Subshell population probabilities $P(M_n)$ for the $2_1^+$ excitation of $^{12}$C by 70 MeV electrons as a function of scattering angle $\vartheta_f$.
$-\cdot -\cdot -$, $M_n=0$; ----------, $M_n=1$; $\cdots\cdots$, $M_n=2$; $-\cdots -\cdots -$, $M_n=-1;\;\,-----,\;M_n=-2$.
The probabilities sum up to unity.}
\end{figure}

The $M_n$-sublevel populations for a collision energy of 70 MeV as a function of scattering angle $\vartheta_f$ are displayed in Fig.2.
It is seen that at scattering angles  below $40^\circ$ all levels have  similar occupation probabilities, in particular the pairs $(+M_n,-M_n)$.
However, in the backward hemisphere, it is just $M_n=0$ and $M_n=1$ which remain important, $M_n=1$ taking over for $\vartheta_f \to 180^\circ$. 
This is due to the strong influence of the magnetic transitions at small electron--nucleus distances (corresponding to a large momentum transfer, respectively to large scattering angles).

\subsection{Decay of the excited nucleus}

In all subsequent results, a coplanar geometry is chosen, where the photon is emitted in the scattering plane,
spanned by $\bfk_i$ 
(which is taken as $z$-axis)  and $\bfk_f$. Thus the azimuthal angle $\varphi$ between
$\bfk_i$ and $\bfk_f$ is $0^\circ$ or $180^\circ$.

For the carbon $2_1^+$ state at $E_x=4.439$ MeV, the ground-state decay width is $\Gamma_n=(1.08\pm0.06)\times 10^{-2}$ eV \cite{AS80}.
The ground-state decay amplitude $A_{fn}^{\rm dec}$ is mediated solely by the current transition densities $J_{L,L\pm 1}$ from Fig.1b,c.
According to the different occupation probabilities of the $M_n$-substates, the intensity of the emitted decay photons depends  on $M_n$ as well.
Fig.3 shows the triply differential cross section for the excitation of the  $M_n$-subshell and its subsequent decay, defined according to (\ref{2.1}) by
$$\frac{d^3\sigma}{d\omega d\Omega_k d\Omega_f}(M_n)\;=\;\frac{2\pi^2\omega^2 E_iE_fk_f}{k_i c^7\;f_{\rm rec}}\;\sum_{\sigma_i,\sigma_f}$$
\begin{equation}\label{2.7}
\times \;\sum_\lambda \left| M_{fi}^{(1)}(M_n)\right|^2,
\end{equation}
where $M_{fi}^{(1)}(M_n)$ is obtained from (\ref{2.2}) if the sum over $M_n$ is dropped, corresponding to the excitation of just one substate $M_n$.
In that case, the photon angular distribution is symmetric with respect to $\theta_k=180^\circ$ and is a superposition of dipole and quadrupole patterns \cite{JP17}.
This symmetry is lost in the total cross section where all $M_n$ subshells are added coherently. In particular, there are angles $\theta_k$ where the total cross section is well below any $M_n$-subshell cross section (for a scattering angle of $\vartheta_f=80^\circ$ near e.g. $\theta_k=40^\circ$, see Fig.3a).
At backward angles (Fig.3b), the total cross section is mainly composed of the $M_n=0$ and $M_n=1$ contributions according to the occupation probabilities from Fig.2.

\begin{figure}

\vspace*{-1.3cm}

\includegraphics[width=10cm]{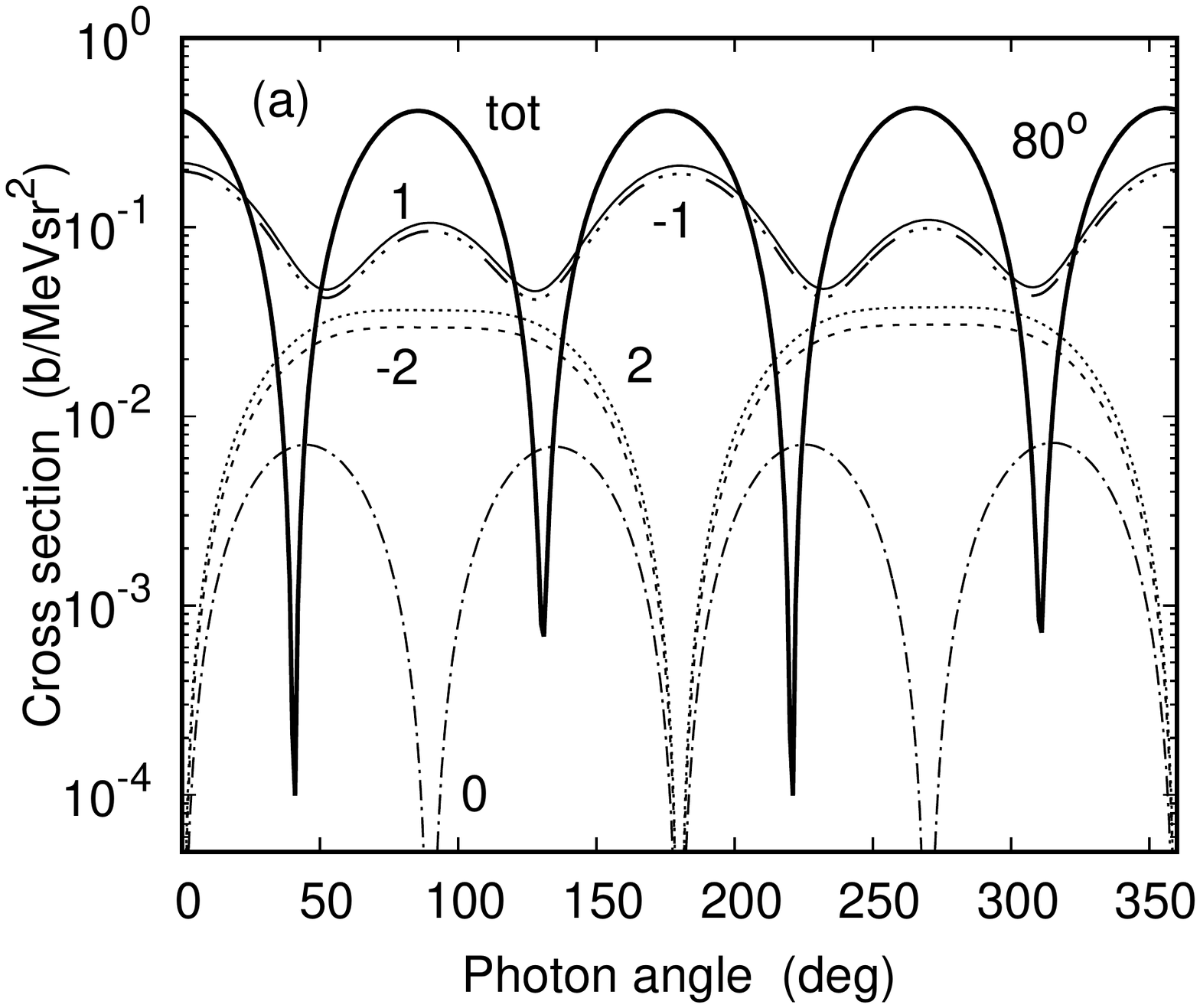}

\vspace*{-1.8cm}

\includegraphics[width=10cm]{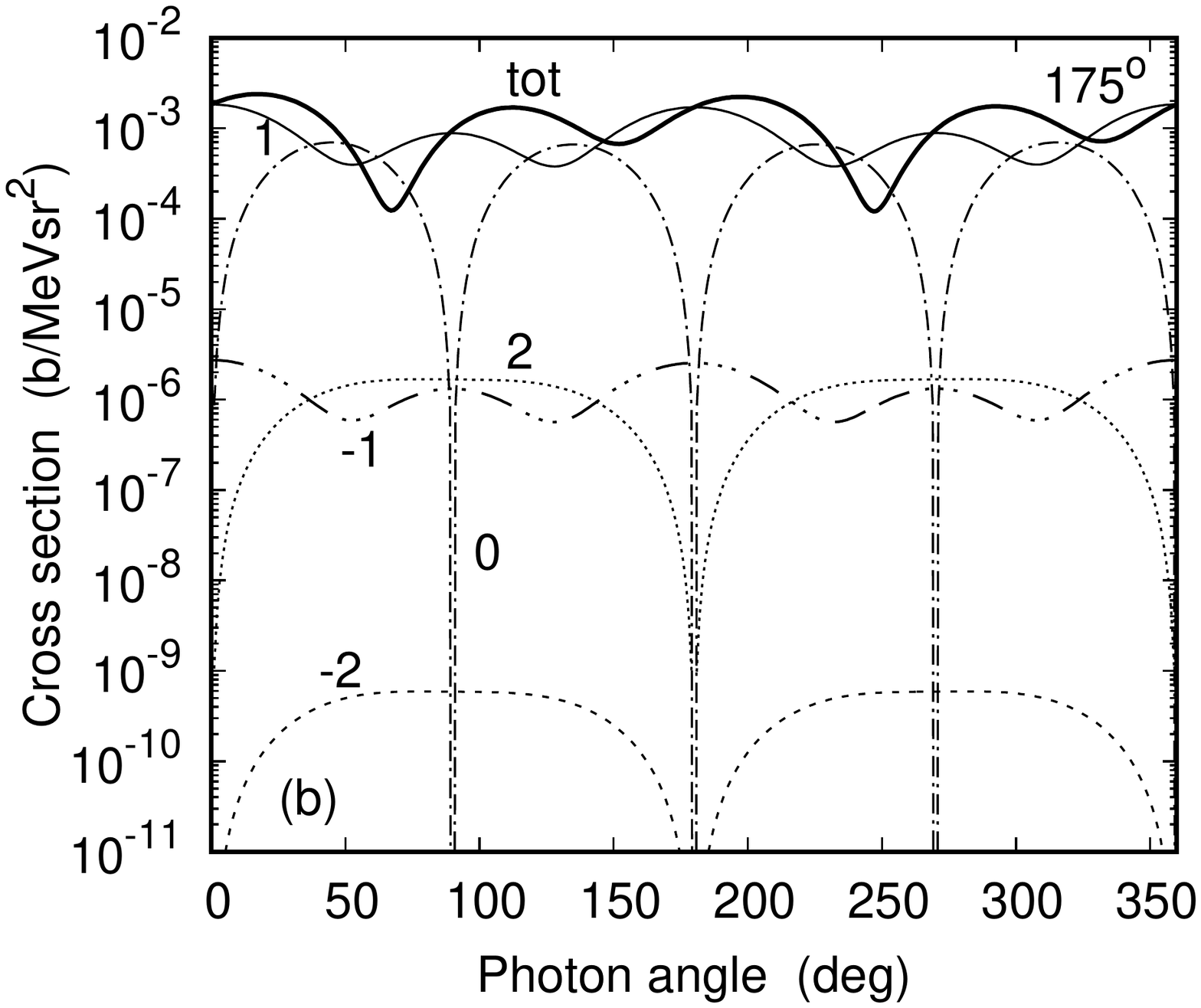}

\vspace*{-0.5cm}

\caption
{$M_n$-subshell cross sections $\frac{d^3\sigma}{d\omega d\Omega_k d\Omega_f}(M_n)$ for the $2_1^+$ excitation of $^{12}$C by 70 MeV electrons and subsequent decay for scattering angles (a) $\vartheta_f = 80^\circ$ and (b) $\vartheta_f=175^\circ$,
with azimuthal angle $\varphi=0$ between electron and photon, as a function of photon angle $\theta_k$.
$-\cdot - \cdot -,\;\,M_n=0;$ -------------, $M_n=1;\;\,\cdots\cdots,\;M_n=2;\;\,-\cdots -\cdots -,\;M_n=-1;\;\,-----,\;M_n=-2.$
Also shown is their coherent sum, the total cross section (thick solid line).}
\end{figure}

In order to display the importance of electric and magnetic excitation, Fig.4 shows the contributions from potential scattering (arising from $\varrho_2$) and from magnetic scattering (due to $J_{23},\,J_{21}$) entering into the excitation amplitude $A_{ni}^{\rm exc}$. Of course, the decay amplitude $A_{fn}^{\rm dec}$
 is kept unchanged in both cases.
In the forward hemisphere, even up to scattering angles $\vartheta_f \sim 160^\circ$, the excitation by the electric force is largely dominant
 at all photon angles (Fig.4a). Only a little shift of the minima to smaller $\theta_k$ is observed when the excitation by the current interaction is included.
Coulomb distortion effects, measured by means of the difference between the DWBA and the PWBA results, can basically be neglected for light nuclei such as $^{12}$C
for not too large scattering angles.
We also note that at $140^\circ$ the photon angular distribution has still the same regular quadrupole pattern as for the smaller angle $80^\circ$ from Fig.3a.
At the backmost angles (Fig.4b for $175^\circ$), the magnetic scattering gives an essential contribution to the cross section, which modulates the quadrupole pattern considerably. This leads to a shift of the minima by about $20^\circ$ as compared to  potential scattering.
Also the Coulomb distortion effects are considerably larger, up to 10 percent.

\begin{figure}

\vspace*{-1.3cm}

\includegraphics[width=10cm]{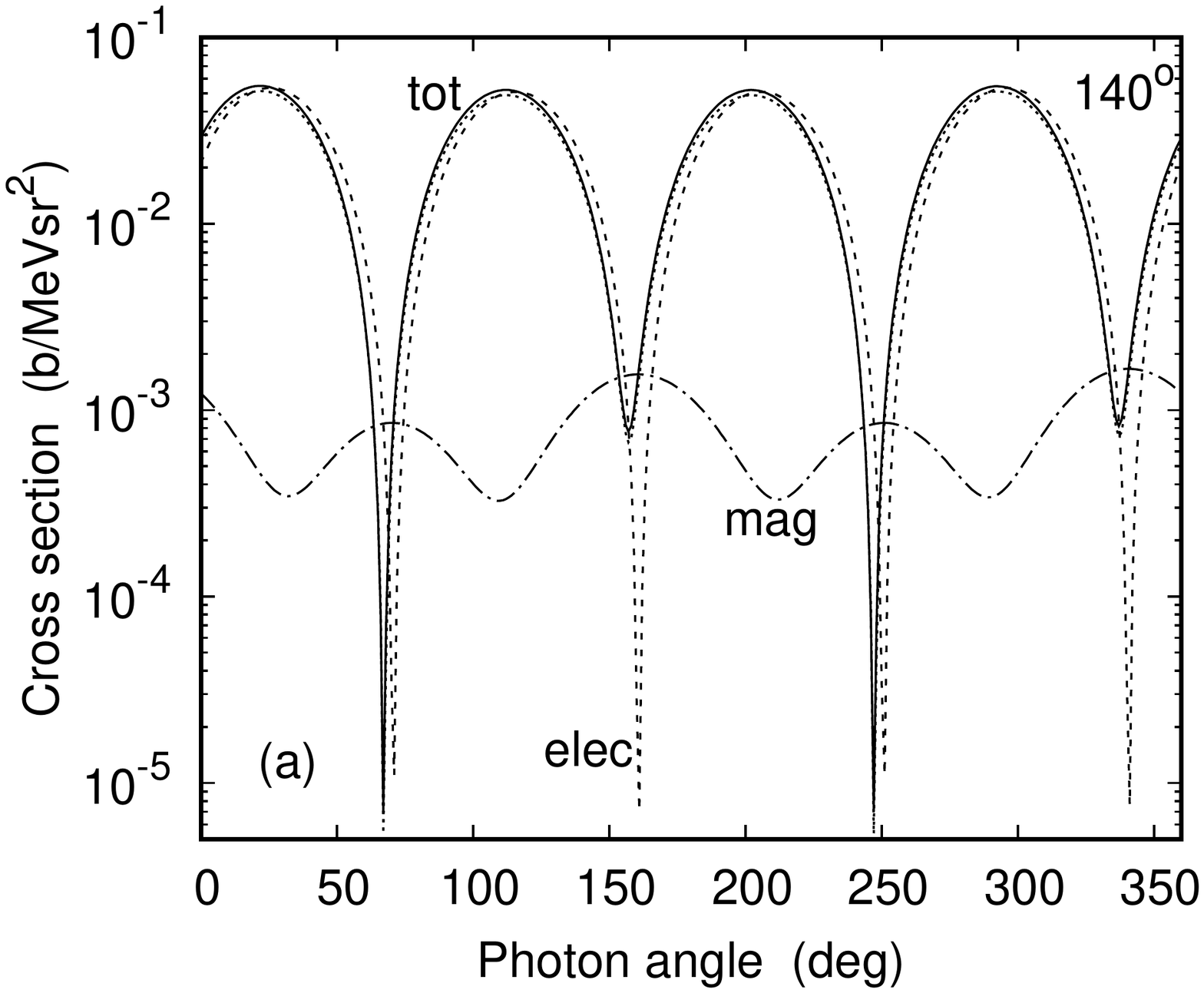}

\vspace*{-1.8cm}

\includegraphics[width=10cm]{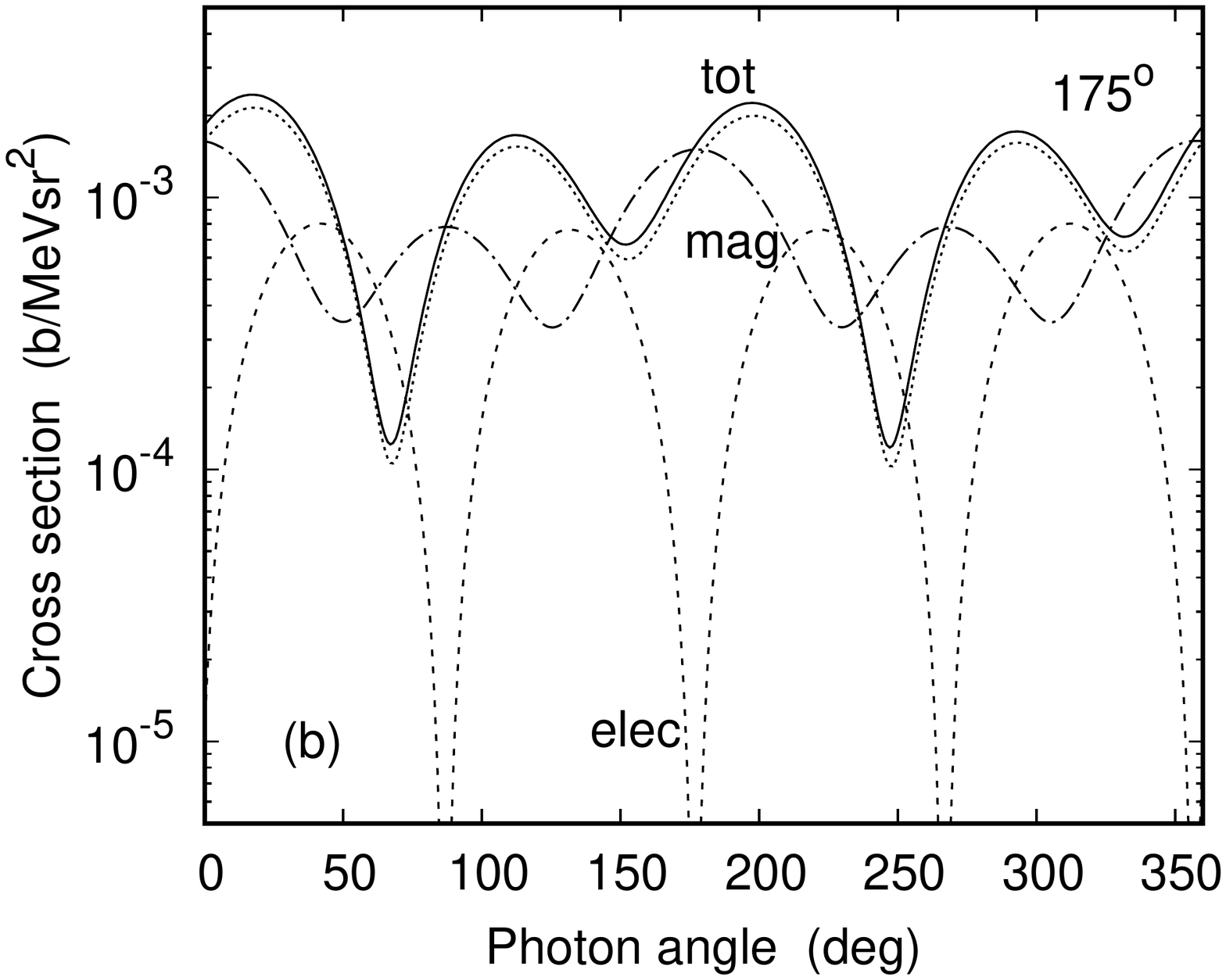}

\vspace*{-.5cm}

\caption
{Triply differential cross section for the nuclear ExDec process of the $^{12}$C, $2_1^+$ state by 70 MeV electrons (a) at $\vartheta_f=140^\circ$ and (b) at $\vartheta_f=175^\circ$, with $\varphi=0$, as a function of photon angle $\theta_k$.
Total cross section:
---------------, DWBA; $\cdots\cdots$, PWBA.
Also shown are the DWBA electric contribution ($-----$) and magnetic contribution $(-\cdot - \cdot -)$ to the total cross section.}
\end{figure}

In Fig.5a the DWBA results from the QPM densities are compared with those based on the Ravenhall et al \cite{Ra87}
densities included in Fig.1. At 
the parameters of the measurements \cite{Pa85},
a collision energy of 66.9 MeV and a scattering angle of $80^\circ$, 
 the Ravenhall  cross section
is enhanced by a factor of 3.5. This results from the higher  transition density $\varrho_2$, since
electric excitation is dominating at this angle.
Included in the figure are results for  potential
scattering within the PWBA, where one of the minima
in the photon angular distribution coincides with
the angle $\theta_q$ which the momentum transfer $\bfq=\bfk_i-\bfk_f$ forms with the z-axis ($\theta_q=312.5^\circ$). This results in an angular distribution which is azimuthally symmetric with respect to $\theta_q$ \cite{Ra87}.
The shift between the minima of the electric PWBA and the full DWBA  is about $2^\circ$, which is verified by the experimental data \cite{Pa85}. These data are measured on a relative scale and are in Fig.5a normalized to the respective theories. It follows from Fig.4a that the shift in angle is basically due to magnetic scattering and not to distortion effects.
In Fig.5b the scattering angle is increased to $170^\circ$. At this angle, the maxima of the Ravenhall results do not coincide anymore with those from the QPM densities.
Although the importance of potential scattering has decreased, the Ravenhall cross sections are still a factor of 3.7 above the QPM ones.
This may be caused by an enhanced current $J_{21}$ near the surface of the nucleus.
Note that the nuclear radius of $^{12}$C is $R_N=1.2\;A^{1/3} = 2.75$ fm, which has to be compared to the distance of closest approach during the electron-nucleus encounter,  determined by the inverse momentum transfer
($q^{-1}= 1.53$ fm for 66.9 MeV and $170^\circ$).

\begin{figure}

\vspace*{-1.3cm}

\includegraphics[width=10cm]{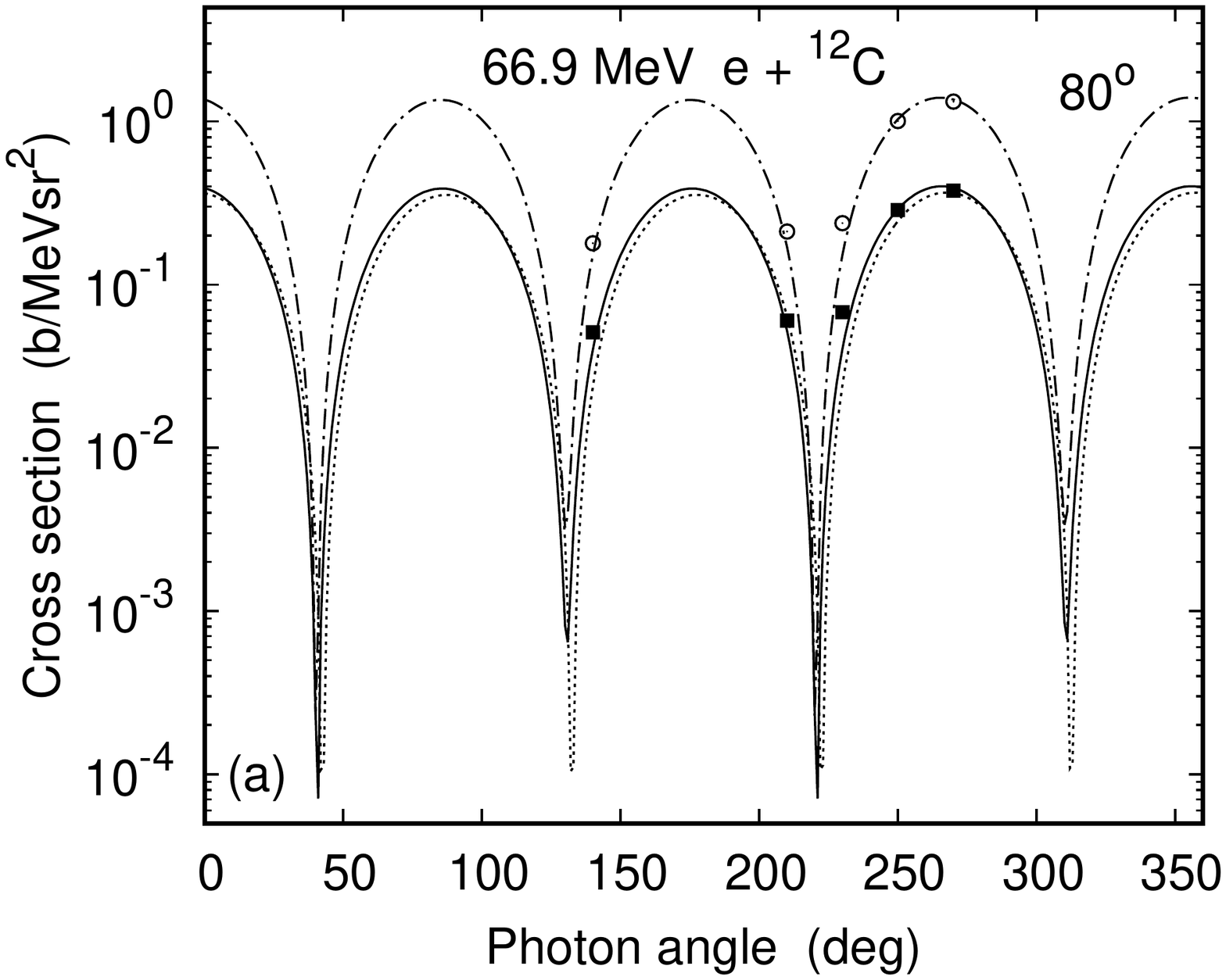}

\vspace*{-1.8cm}

\includegraphics[width=10cm]{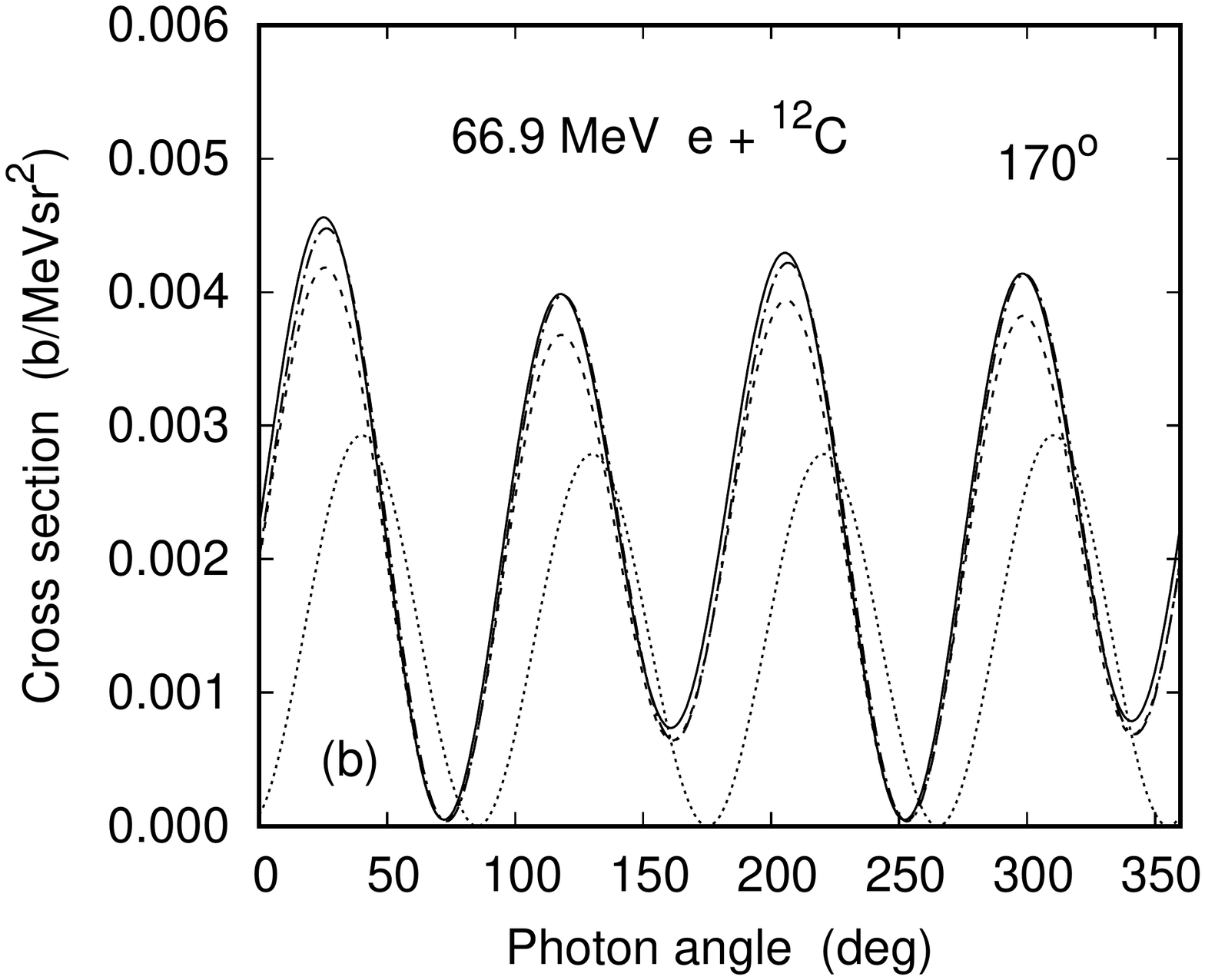}

\vspace*{-.5cm}

\caption
{Triply differential cross section for the nuclear ExDec process of the $^{12}$C,$\;2_1^+$ state by 66.9 MeV electron impact at (a) $\vartheta_f=80^\circ$ and (b) $\vartheta_f=170^\circ$, with $\varphi=0$,  as a function of photon angle $\theta_k$.
-----------------, DWBA results; $-----$, PWBA results (in (b)); $\cdots\cdots$, PWBA results for electric excitation (all with QPM densities);
$-\cdot - \cdot -$, DWBA results with Ravenhall densities. In (b), the Ravenhall results are scaled down by a factor of 0.27
to display the differences in shape.
The experimental data $(\circ)$ in (a) are taken from Papanicolas et al \cite{Pa85}.
The same data $(\blacksquare)$ are scaled down by a factor of 0.285 to fit the QPM results.}
\end{figure}

\subsection{Influence of bremsstrahlung}

In order to give predictions for the contribution of brems\-strah\-lung to the nuclear ExDec process it is important to account for the finite  resolution $\Delta \omega$ of the photon detector.
As far as the nuclear ExDec process with its resonant behaviour is concerned, the averaging over the detector resolution leads basically to a reduction of intensity, but not to a change in the photon angular distribution.
Bremsstrahlung, on the other hand, due to its weak dependence on $\omega$, is hardly affected by the averaging procedure.
When both contributions are considered, the averaged photon intensity at the peak frequency $\omega=E_x$ is calculated from
$$\left\langle \frac{d^3\sigma}{d\omega d\Omega_k d\Omega_f}\right\rangle_{\Delta \omega}\;\approx\; \frac{2\pi^2 E_i}{k_ic^7}\sum_{\sigma_i,\sigma_f}\;\sum_\lambda$$
\begin{equation}\label{2.8}
\times \;\frac{1}{\Delta \omega}\int_{E_x-\frac{\Delta \omega}{2}}^{E_x+\frac{\Delta \omega}{2}} d\omega'\;\frac{\omega^{'2} E_fk_f}{f_{\rm rec}}\;\left| M_{fi}^{(1)}\;+\;M_{fi}^{\rm brems}\right|^2,
\end{equation}
such that the different $\omega$-behaviour of $M_{fi}^{(1)}$ and $M_{fi}^{\rm brems}$ leads to a change in the $\theta_k$-distribution which strongly depends on $\Delta \omega$.
This feature is displayed in Fig.6a where a resolution of $\Delta \omega/\omega = 3\%$ is taken, corresponding to a LaBr photon detector to be used in experiments,
while in Fig.6b, $\Delta \omega /\omega = 0.5\%$ is assumed.
Again, the experimental parameters, $E_e=E_i-c^2=66.9$ MeV,  the scattering angle $\vartheta_f= 80^\circ$ and $\omega=4.439$ MeV, have been chosen.
The bremsstrahlung angular distribution is characterized by the narrow double-peak structure near $\theta_k=0$ and near $\theta_k=\vartheta_f$ for $\omega \ll E_e$.
These structures dominate the photon distribution from the ExDec process. In addition, the bremsstrahlung photons fill the minima of the quadrupole pattern, the more so, the poorer the detector resolution.
We note that the experimental data points are slightly better reproduced with a resolution near or below 1\%.

\begin{figure}

\vspace*{-1.3cm}

\includegraphics[width=10cm]{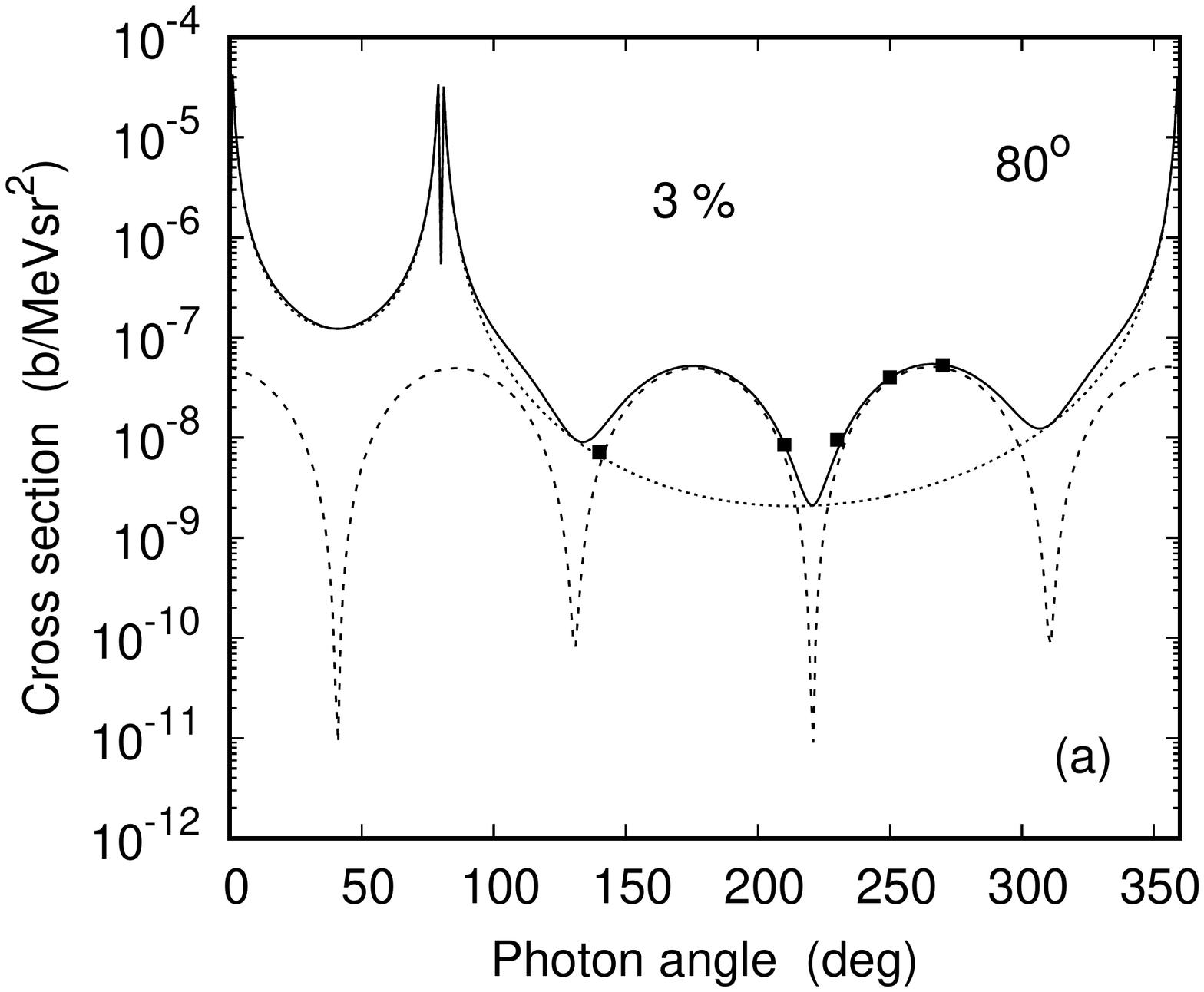}

\vspace*{-1.8cm}

\includegraphics[width=10cm]{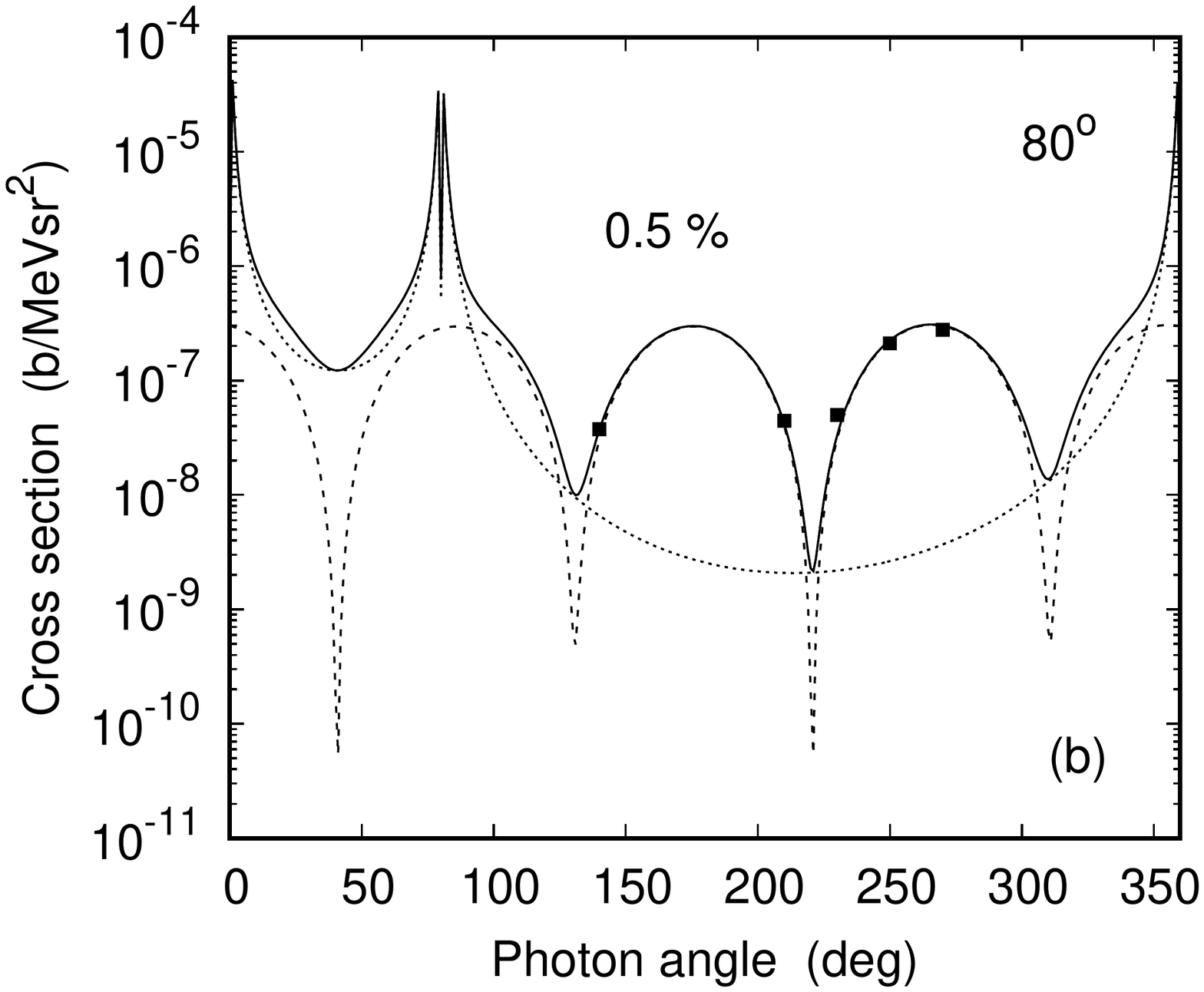}

\vspace*{-.5cm}

\caption
{Averaged triply differential cross section for the coincident $(e,e'\gamma)$ process by 66.9 MeV electrons scattered at $\vartheta_f=80^\circ$, with $\varphi=0$,  as a function of photon angle $\theta_k$.
The detector resolution is (a) $\Delta \omega /\omega =3\%$ (corresponding to $\Delta \omega =133$ keV), and (b) $\Delta \omega/\omega=0.5\%$ (corresponding to $\Delta \omega = 22.2$ keV).
$-----$, photons from the nuclear ExDec process; $\cdots\cdots$, bremsstrahlung; ---------------, coherent sum.
The  experimental data $(\blacksquare)$ are from Papanicolas et al \cite{Pa85} and are normalized to the full lines.}
\end{figure}

In order to study the influence of bremsstrahlung at other geometries we display in Fig.7 photon angular distributions at two scattering angles in the backward hemisphere, $\vartheta_f=140^\circ$ and $170^\circ$, and two collision energies, 70 MeV and 150 MeV.
In all subfigures, an average is taken with $\Delta \omega / \omega =3\%$.
Comparing Figs.6a, 7a and 7b, it is seen that, away from the bremsstrahlung peaks, the influence of bremsstrahlung decreases with scattering angle, favouring the backmost angles.
Also, profiting from a weak dependence of the nuclear ExDec process on collision energy \cite{JP17}, while bremsstrahlung is strongly decreasing with $E_e$, bremsstrahlung is
the more suppressed, the higher $E_e$, see Figs.7c and 7d.

\begin{figure}

\vspace*{-1.3cm}

\includegraphics[width=10cm]{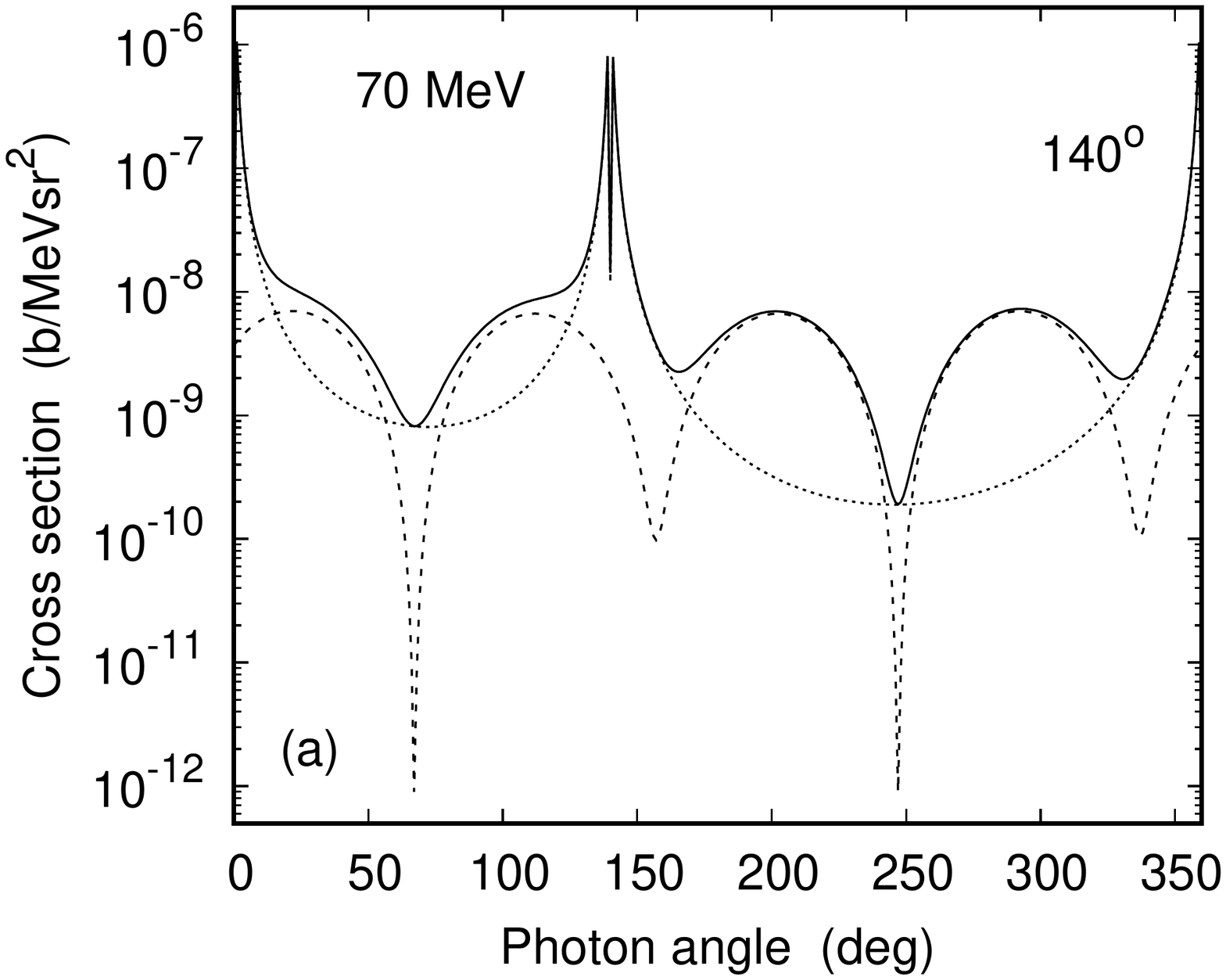}

\vspace*{-1.8cm}

\includegraphics[width=10cm]{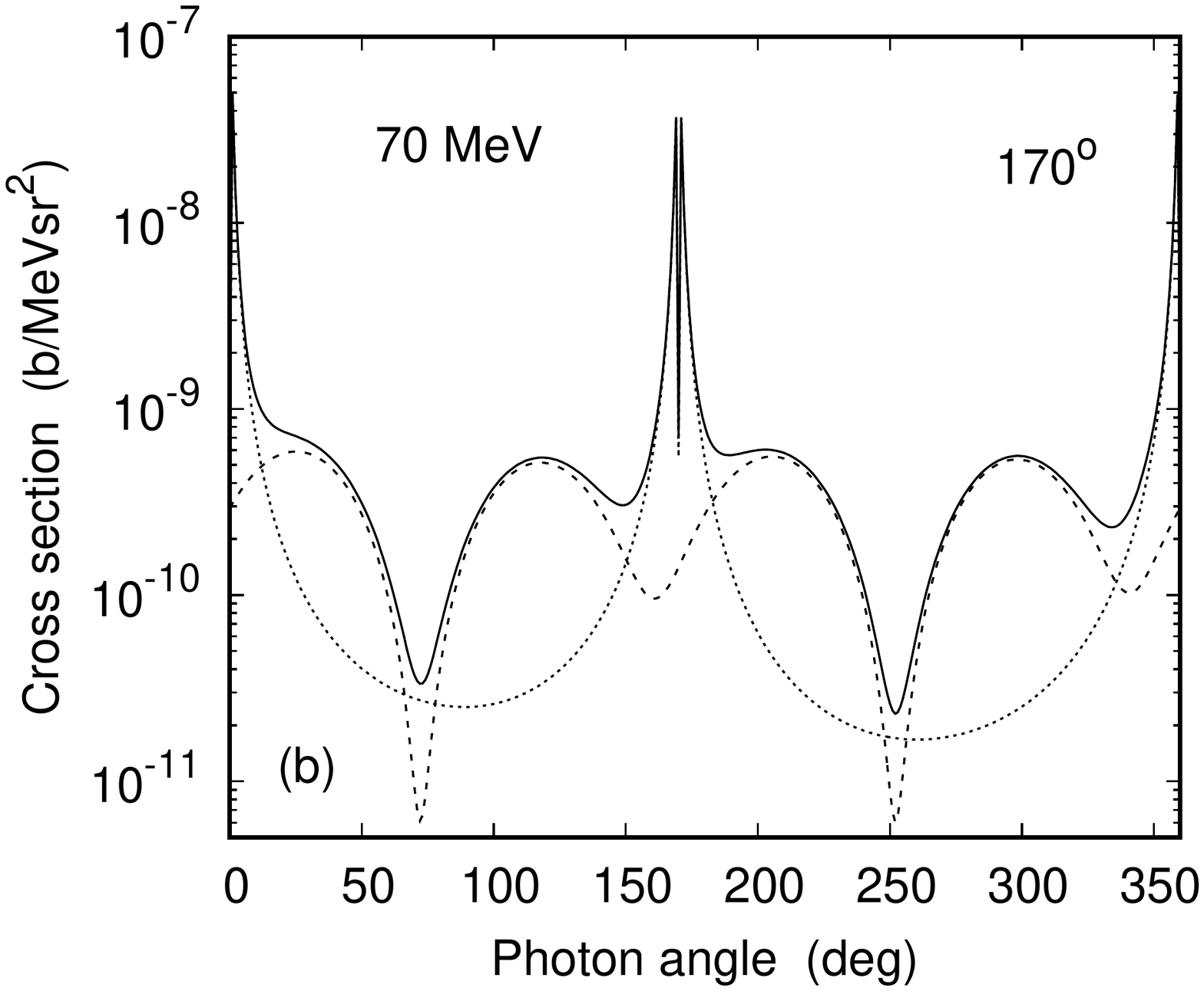}

\vspace*{-1.8cm}

\includegraphics[width=10cm]{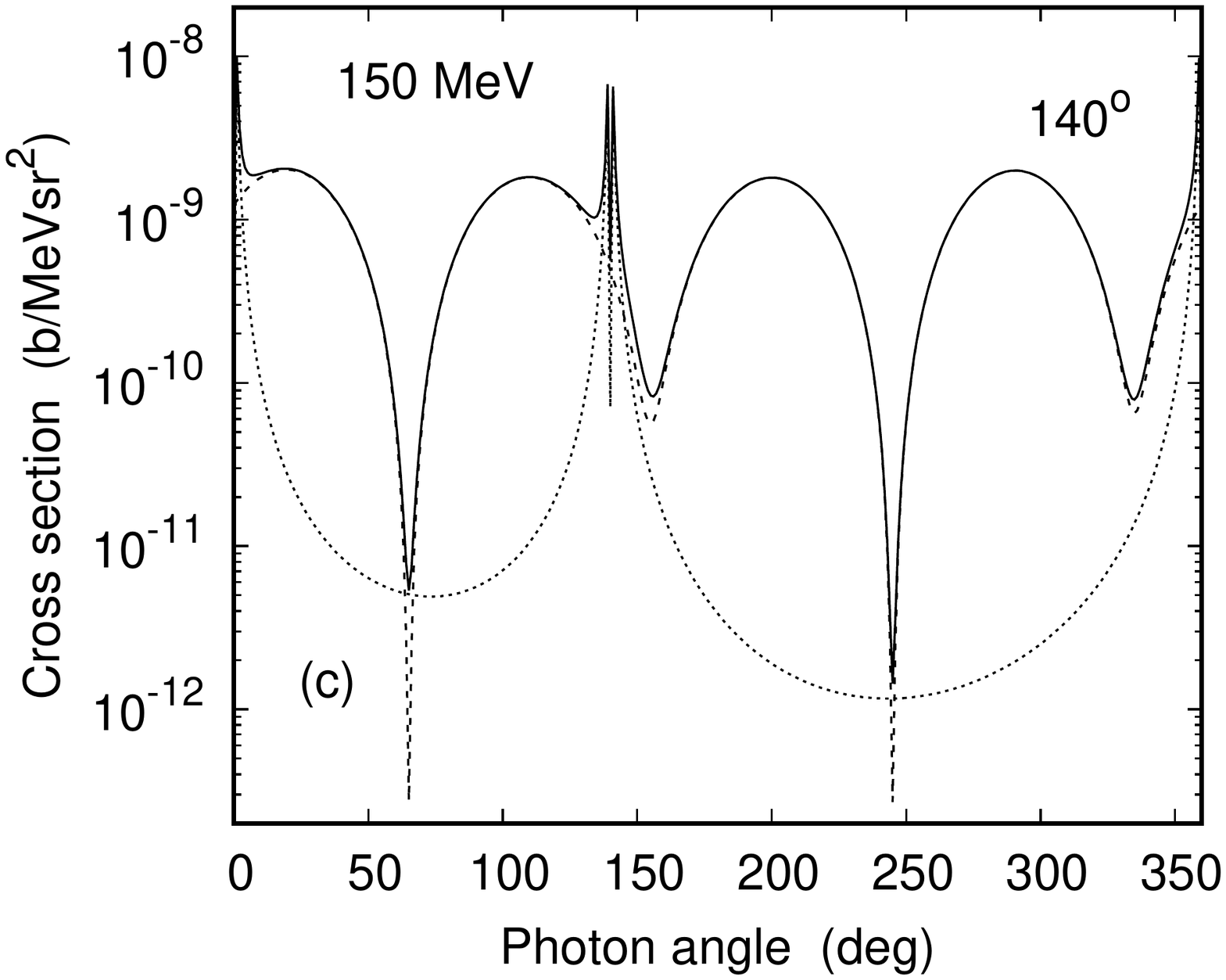}

\vspace*{-1.8cm}

\includegraphics[width=10cm]{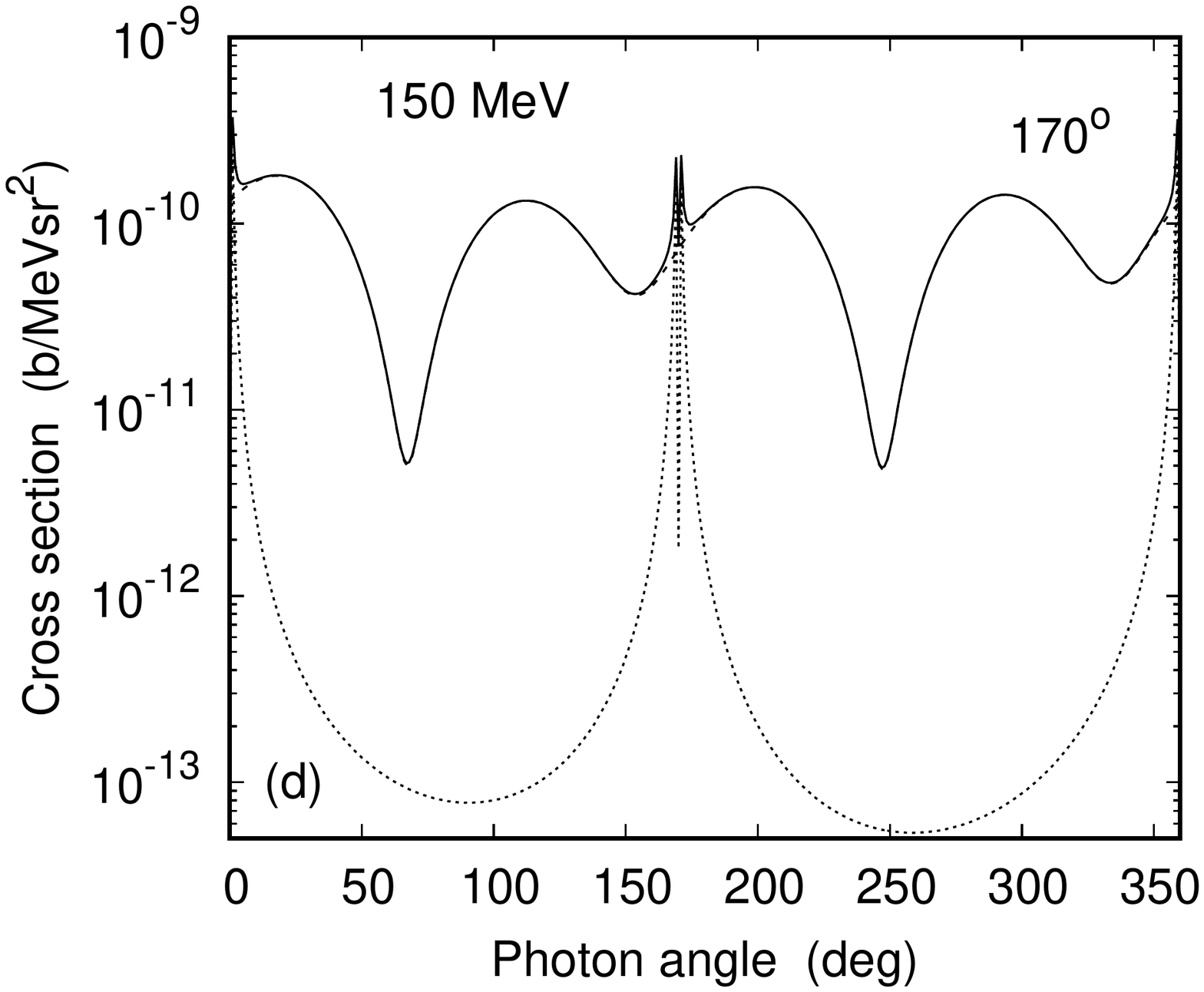}

\vspace*{-.5cm}

\caption
{Averaged triply differential cross section for the coincident $(e,e'\gamma)$ process by (a), (b) 70 MeV and (c), (d) 150 MeV electrons as a function of photon angle $\theta_k$.
The scattering angle is (a), (c) $\vartheta_f=140^\circ$ and (b), (d) $\vartheta_f=170^\circ$, at $\varphi=0$.
The detector resolution is $\Delta \omega/\omega =3\%$.
$-----$, photons from the nuclear ExDec process; $\cdots\cdots$, bremsstrahlung; ----------------, coherent sum.}
\end{figure}

\section{Polarized electrons}
\setcounter{equation}{0}

Previous investigations of the ExDec process were restric\-ted to unpolarized beam electrons. However, a more stringent test of the nuclear models is achieved if additionally 
the spin degrees of freedom are taken into account. 
An appropriate measure of the spin asymmetry is the Sherman function $S$ \cite{Sh56,Mo}  which requires a beam polarization perpendicular
to the scattering plane.
It measures the relative difference in intensity when the direction of the beam polarization is switched.

In the discussion of the spin asymmetry we will disregard bremsstrahlung, since actual measurements will always be performed at photon angles where the influence of bremsstrahlung is small. In that case, the Sherman function can alternatively be obtained from the transition amplitude $M_{fi}^{(1)}$. Denoting the coefficients of the initial-state polarization vector $\bfzeta_i$ in the standard 
basis ${1 \choose 0}$ and ${0 \choose 1}$ by $a_{m_i},\;M_{fi}^{(1)}$ is formally written as \cite{Ros}
\begin{equation}\label{3.1}
M_{fi}^{(1)}\;=\;\sum_{m_i=\pm \frac12} a_{m_i}\;F(m_i),
\end{equation}
and the Sherman function results from \cite{Jaku15}
\begin{equation}\label{3.2}
S\;=\;-\;2\;\frac{\sum_\lambda \mbox{Im }\{F^\ast(\frac12)\cdot F(-\frac12)\}}{\sum_\lambda \left[ \,|F(\frac12)|^2\,+\,|F(-\frac12)|^2\right]}.
\end{equation}
The denominator is proportional to the total cross section for unpolarized particles, obtained by summing (in addition to $m_i$)
over the two photon polarizations $\bfeps_\lambda$ and over the projections of the  two spin polarization vectors $\bfzeta_f$ of the final electron (note, however,  that
$F(m_i)$ is independent of $\sigma_f$
 if $\bfzeta_f$ is taken parallel, respectively, antiparallel to $\bfk_f$).

\begin{figure}

\vspace*{-1.3cm}

\includegraphics[width=10cm]{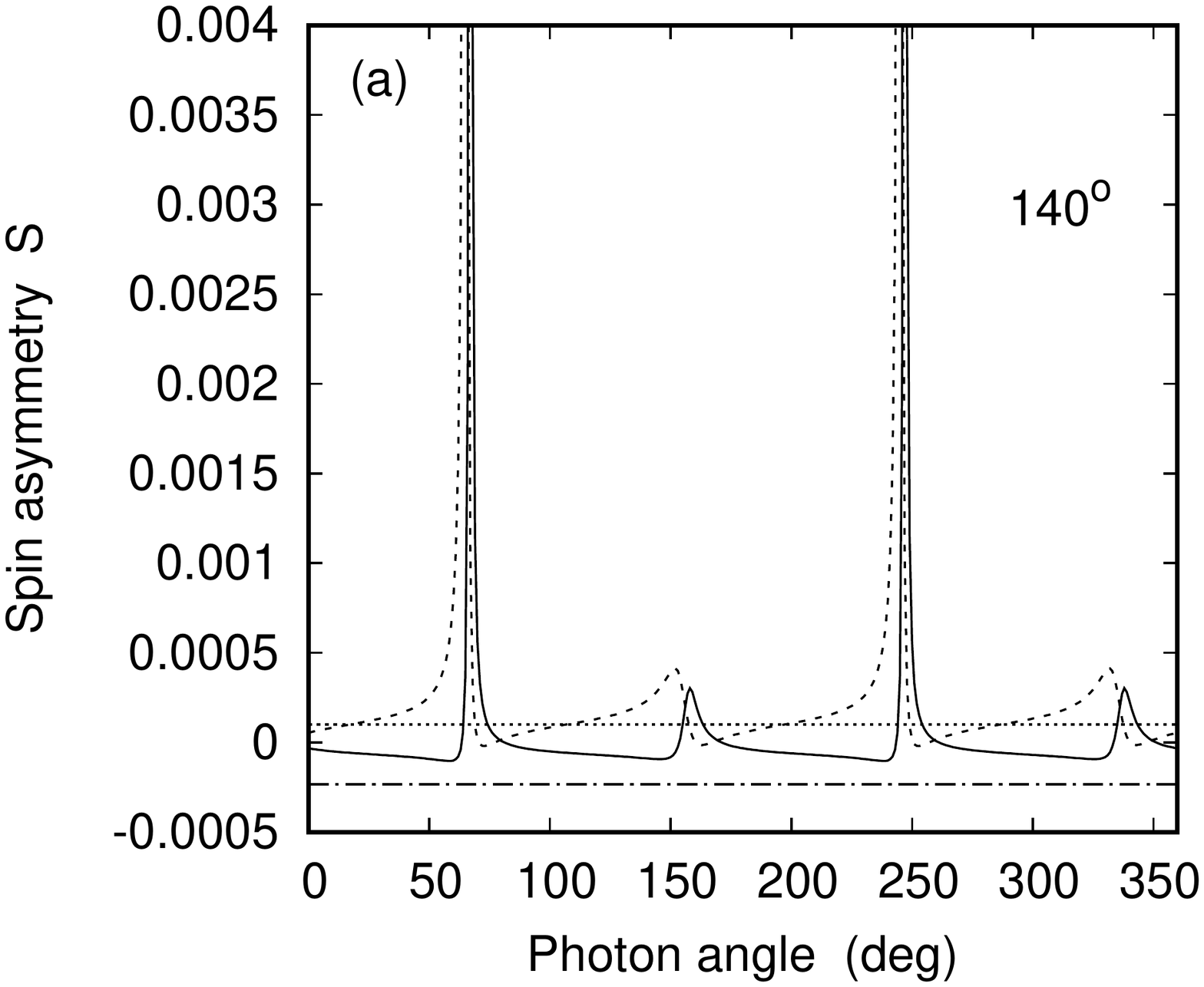}

\vspace*{-1.8cm}

\includegraphics[width=10cm]{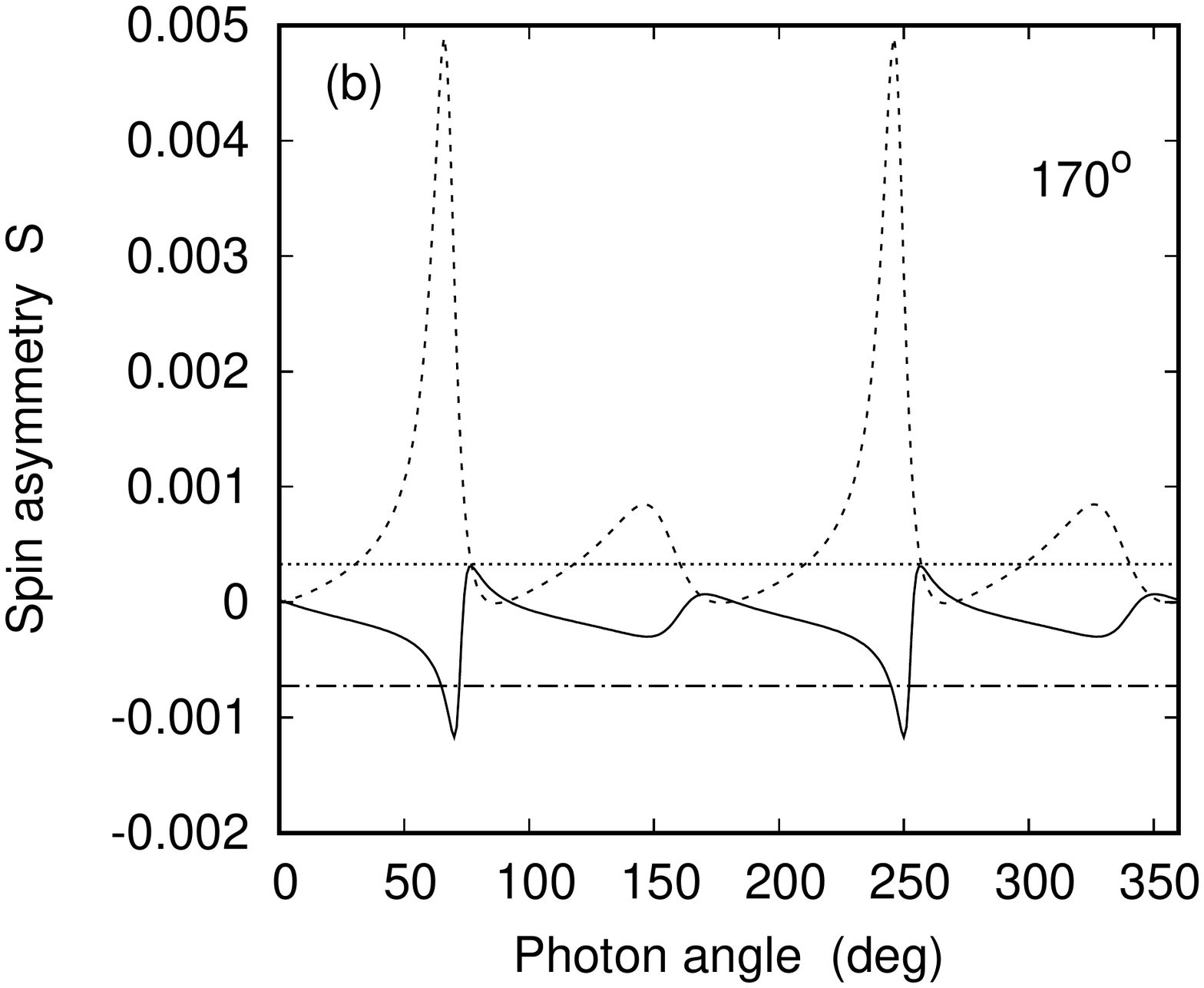}

\vspace*{-.5cm}

\caption
{Spin asymmetry for the nuclear ExDec process from the carbon $2_1^+$ excitation by 70 MeV (---------------) and 150 MeV $(-----)$ perpendicularly polarized electrons as a function of photon angle $\theta_k$ for scattering angles (a) $\vartheta_f=140^\circ$ and (b) $\vartheta_f=170^\circ$  at $\varphi=0.$
The maxima of $S$  in (a) amount to 0.031 for 70 MeV and to 0.066 for 150 MeV (using a step size of $\Delta \theta_k=1^\circ$ in the plot).
Also shown is the spin asymmetry from the excitation process alone ($-\cdot -\cdot -,$ 70 MeV; $\cdots\cdots,$ 150 MeV).} 
\end{figure}

Fig.8 provides  examples for the spin asymmetry in case of some geometries from Fig.7.
The total cross section being in the denominator of (\ref{3.2}), $S$ has extrema at photon angles where the minima of the cross section are located (see also Fig.9).
In the forward hemisphere, and even at scattering angles up to $140^\circ$, the cross section has very deep minima and consequently, the maxima of $S$ are very sharp.
In a true experimental situation the excursions of $S$ at such angles will be reduced since bremsstrahlung tends to fill the cross section minima.
At the backmost scattering angles, diffraction effects come into play and modulate the sign of $S$.
Such diffraction effects occur when the electron is sufficiently energetic to penetrate the nuclear surface and to scatter off the individual protons. 

In Fig.8 we have included the spin asymmetry resulting from the mere excitation process as horizontal lines.
It is calculated by replacing $M_{fi}^{(1)}$ with the amplitude $A_{ni}^{\rm exc}$, for which an equation of type (\ref{3.1}) also holds. In (\ref{3.2}), the sum over $\lambda$ has to be changed into a sum over $M_n$ \cite{Jaku15}.
For excitation it is well known (and verified in Fig.8) that $|S|$ decreases globally with $E_i$ (at fixed $\vartheta_f$) and increases with scattering angle (at fixed $E_i$). It is only the latter fact which remains true for the nuclear ExDec process. 

\begin{figure}[t]

\vspace*{-1.3cm}

\includegraphics[width=10cm]{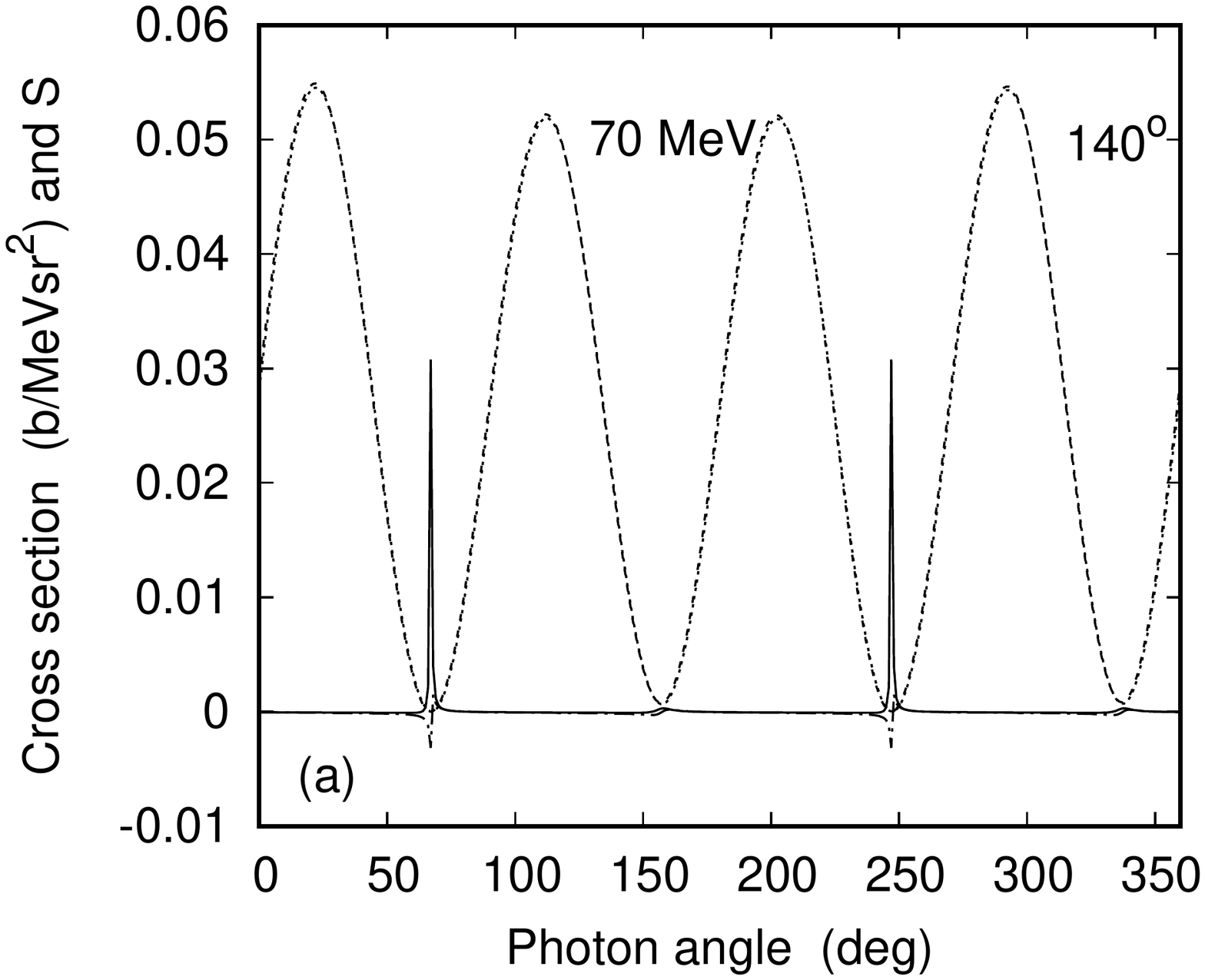}

\vspace*{-1.8cm}

\includegraphics[width=10cm]{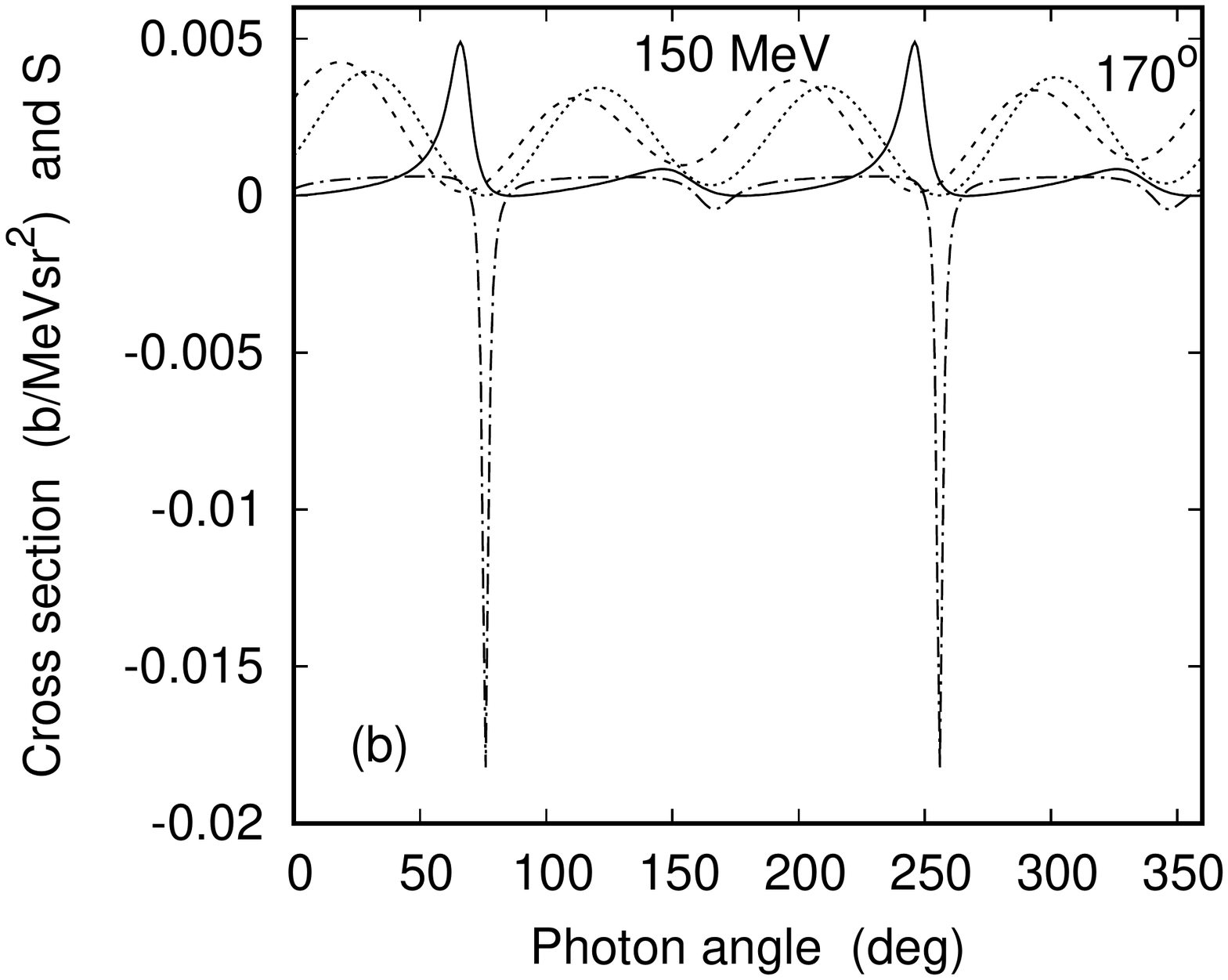}

\vspace*{-.5cm}

\caption
{Spin asymmetry for nuclear excitation and decay of the $^{12}$C, $2_1^+$ state by  perpendicularly polarized electrons of (a) 70 MeV 
at $\vartheta_f=140^\circ$ and (b) 150 MeV  at $\vartheta_f=170^\circ$, with $\varphi=0$,  as a function of photon angle $\theta_k$.
Shown is $S$ from the nuclear ExDec process using the QPM densities (---------------) and the Ravenhall densities ($-\cdot - \cdot -$).
Also shown is the corresponding (unaveraged) triply differential cross section (in $\frac{b}{MeVsr^2}$) from the QPM densities $(-----)$ and from the Ravenhall densities $(\cdots\cdots)$.
In (a), the Ravenhall cross section is scaled down by a factor of 0.26, in (b) the QPM cross section is scaled up by a factor of 3 to display the differences in shape.}
\end{figure}

In order to demonstrate the greater sensitivity of $S$ to the choice of nuclear models as compared to the perceptivity of the cross section for unpolarized particles, we display in Fig.9 the results obtained from the QPM transition densities on the one hand, and from the Ravenhall transition densities on the other hand.
At a beam  energy of 70 MeV and $\vartheta_f=140^\circ$ (Fig.9a)
the Ravenhall cross section is by a factor of 3.85 higher, but the shape of the angular distribution is nearly
identical. The Sherman function, however, differs visibly. In particular, the maxima in $S$ from the QPM prescription have turned into weak minima in the Ravenhall picture. 
The sensitivity to details in the transition densities increases with energy.
In Fig.9b a collision energy of 150 MeV is chosen, together 
with a backward scattering angle of $170^\circ$ which increases the spin asymmetry in the regions between the sharp peaks considerably.
In this geometry, the Ravenhall cross section is enhanced by a factor of 3, and the angular distribution is slightly modulated and shifted.
The Sherman function, on the other hand, shows considerable deviations in the two prescriptions. The maxima of the QPM model have now turned into deep minima. These extrema are nevertheless  wide enough to make a detection feasible. Moreover, as becomes clear from a comparison with Fig.7d,
bremsstrahlung plays no role except in a small region around $\theta_k=170^\circ$, so that the extrema in $S$ are not influenced.

\section{Conclusion}

We have calculated the triply differential cross section for the simultaneous observation of the scattered electron and the emitted photon in the $(e,e'\gamma)^{12}$C reaction. The nuclear quasiparticle phonon model 
was used for the  excitation of the $2_1^+$ state, while electron scattering was described within the distorted-wave Born approximation.
Comparing with earlier results using experimental nuclear transition densities, large changes in the  photon intensity are found, but only slight shifts of the angular distribution, even at backward scattering angles.
The measured relative photon distribution is well reproduced in both prescriptions.

Confirming earlier results on quadrupole excitation of $^{92}$Zr, the $M_n$-sublevels of the $^{12}$C, $2_1^+$ excited state are approximately equally populated for scattering angles in the forward hemisphere, while the $M_n=0$ and $M_n=1$ substates largely
dominate at the backmost angles. Consequently, at the smaller angles the photon angular distribution has a regular quadrupole structure, while there are substantial dipole-type modifications (from the $M_n=1$ contribution) at scattering angles close to $180^\circ$.

Including bremsstrahlung within the PWBA, a theory well justified for low-energy photons and a light nucleus like $^{12}$C even for large scattering angles, it was found that for photon angles in the forward direction or close to the scattering angle, bremsstrahlung spoils the visibility of the nuclear decay photons, the more so, the smaller the scattering angle, the lower the collision energy and the poorer the resolution of the photon detector.

Finally we have investigated the Sherman function which is a measure of the spin asymmetry occurring for polarized electron impact.
In contrast to its behaviour for elastic scattering or excitation where the spin asymmetry exhibits a global decrease with collision energy (which may be modulated by diffraction structures),
the ExDec process will lead to considerably higher spin asymmetries when $E_e$ is increased.
Furthermore, by comparing the results from the  two considered types of nuclear transition densities, we have  demonstrated that the Sherman function is much more sensitive to such changes than the triply differential cross section.
The large deviations of $S$ in the two models at high collision energies, combined with its high absolute values at the backmost scattering angles where the  influence of bremsstrahlung is negligible,
make such a geometry a promising candidate for nuclear structure investigations.

V.Yu.~P. acknowledges support by the Deutsche For\-schungs\-gemeinschaft 
(DFG, German Research Foundation) - Projektnummer 279384907 - SFB 1245.




\vspace{1cm}

\end{document}